\documentclass[twocolumn, amsmath,amssymb,amssymb,aps, showpacs,pra]{revtex4}
\usepackage{amssymb}
\usepackage{amsmath}
\usepackage{graphicx}
\usepackage{dcolumn}
\usepackage{bm}
\usepackage{braket}
\usepackage{mathrsfs}
\usepackage{pxfonts}
\usepackage{graphicx}
\begin{document}
\title{Using Quantum Coherence to Enhance Gain in Atomic Physics}
\author {Pankaj K. Jha\footnote{Present Address: (NSF) Nanoscale Science and Engineering Center(NSEC), 5130 Etcheverry Hall, University of California, Berkeley, California 94720-1740, USA. Electronic Address: pkjha@berkeley.edu}}
\affiliation{Department of Physics and Astronomy, Texas A$\&$M University, College Station, Texas 77843, USA\\
School of Engineering and Applied Science, Princeton University, Princeton, New Jersey 08544, USA }
\date{\today}
\pacs{42.65.-k, 42.65.Ky, 42.50.Lc}    
\begin{abstract}
Quantum coherence and interference effects in atomic and molecular physics has been extensively studied due to intriguing counterintuitive physics and potential important applications. Here we present one such application of using quantum coherence to generate and enhance gain in extreme ultra-violet(XUV)(@58.4nm in Helium) and infra-red(@794.76nm in Rubidium) regime of electromagnetic radiation. We show that using moderate external coherent drive, a substantial enhancement in the energy of the lasing pulse can be achieved under optimal conditions. We also discuss the role of coherence. The present paper is intended to be pedagogical on this subject of coherence-enhanced lasing.
\end{abstract}
\maketitle
\section{Introduction}
Interaction of light with matter is a fundamental areas of research in quantum optics and atomic physics. Quantum coherence and interference\cite{Hanle} has led to many novel effects\cite{MOS} for e.g. coherent population trapping~\cite{CPT,CPT1,CPT2}, amplification or lasing without population inversion(LWI)~\cite{LWI,LWI12,LWI4,LWI5}, ultraslow light~\cite{SL,SL1,SL2}, enhancement of refractive index without absorption~\cite{REE,REE1,REE2,REE3}, highly sensitive magnetometry~\cite{HSM,HSM1}, coherent Raman umklappscattering\cite{CRU}, high resolution nonlinear spectroscopy\cite{Jain93, Jain96}, sensing nanoscale molecular complexes\cite{Voronine12}, photodesorption\cite{JhaAPL12}, bridging quantum optics with position dependent mass Schrodinger equation(PDMSE)\cite{JhaPQE11} etc. Recent experimental and theoretical studies have also provided support for the hypothesis that even biological systems use quantum coherence\cite{Engle07,Collini10,Pani10}. Nearly perfect excitation energy transfer in photosynthesis is an excellent example of this. Furthermore, during the past decade study of quantum interference(QI) effects has been extended to tailored semiconductor nanostructures like quantum wells and dots due to coherent resonant tunneling owing to their potential applications in photo-detection \cite{Woj08, Vas11}, lasing \cite{Cap97,Sch97}, quantum computing and quantum circuitry \cite{And10, Gib11}, optical modulator\cite{Carter}. Generally these coherence are generated using coherent source (laser) to manipulate the optical response of the system. But coherences can also be induced by vacuum also know as vacuum induced coherence (VIC)\cite{Agarwal74}. Interplay between the coherence generated by laser + microwave source has also been studied in regard to microwave controlled electromagnetically induced transparency\cite{Li09}, four-wave mixing\cite{Zibrov02}, Raman and SubRaman generation\cite{JhaPRA13} etc.

Recently quantum coherence effects has extended its domain to plasmonics with a recent proposal of coherence-enhanced spaser\cite{Dorfman13} and propagation of surface plasmon polaritons\cite{PKJha13}. Coherent control in plasmonics will definite add a new dimension to the field of nanophotonics. On one hand quantum coherence effects in quantum optics and atomic physics is a subject of intense theoretical and experimental investigation while on the other hand its effect in human brain has been a topic of debate and discussion\cite{Mavro}.

Coherent excitation in two-level system, studied by Mollow\cite{Mollow69}, brought interesting features in the resonance fluorescence spectrum which was later confirmed by the beautiful experiment\cite{Mollow77}. A counterpart of the Mollow's triplet\cite{Mollow69} was observed with incoherent excitation in a cavity by Valle and Laussy\cite{Valle10} where they showed that the strong-coupling between the cavity and the emitter generates the necessary coherence required. For multi-level system the coherence can be easily generated by coupling the upper-level to an adjacent level with a coherent electromagnetic field. Recently Scully\cite{Scu10} extended the idea of coherence effects to solar photovoltaic cells and showed that such devices can benefit from quantum boost. In fact this coherence in solar cells can be generated by an external source like microwave radiation source or by noise-induced quantum interference which is essentially different from the former which costs energy\cite{Scu11,Jha11}.  

Although numerous theoretical and experimental studies of coherence effects have been performed, there are still open areas to be explored. For example, quantum coherence and interference which plays a key role in LWI as shown extensively in the literature, the burning question we always ask: \textit{Can it be used as a tool for enhancing the gain in the X-Ray/XUV regimes of electromagnetic radiation}? A realistic approach in this area may open a door for the development of more powerful lasers in the wavelength down to  \textquotedblleft water window\textquotedblright\,. One approach was proposed by Scully\cite{mos08jmo} in which it was shown that intense short pulses XUV radiations can be produced by cooperative spontaneous emission or Dicke superradiance\cite{Dicke54} from visible or IR pulses. Later on we also proposed using coherence to generate\cite{TLWI2}  and enhance \cite{PKJha12} gain in XUV regime with Helium, Helium-like Carbon, Boron as our gain medium. A unique way to accomplish effective unidirectional excitation using bi-directional source was discussed in Ref.\cite{PKJha11} to boost gain in the XUV regime.

In this paper we review and extend the approach of applying a strong driving field on an adjacent transition to the lasing transition to enhance gain and show that gain can be substantially (more than an order of magnitude) increased under optimal conditions. Here we have discussed two regimes, \textit{transient} and \textit{steady-state}, for coherence enhanced lasing. We have considered lasing on extreme ultra-violet(XUV) transition of Helium and $\text{D}_{1}$ transition of Rubidium in the transient and the steady-state regimes respectively. 

This paper is organized as follows. In section II, we discuss the inversion requirement for lasing in the two-level emitter based gain medium. In section III we discuss the effect of coherent drive on gain for three-level emitter in $\Lambda$-configuration(see Fig.~\ref{levels}) with initial population inversion($\varrho_{aa}(0)+\varrho_{cc}(0)>\varrho_{bb}(0)$) in the two limits of ratio between the spontaneous decay rate on the drive transition$(\gamma_{c})$ and the lasing transition$(\gamma_{b}), $ (a) $\gamma_{c} \gg \gamma_{b}$ and (b) $\gamma_{c} \ll \gamma_{b}$. In section IV we consider Rubidium laser at $D_{1}$ transition and study the effect of coherent drive on the output energy in the steady-state regime. In section V, we present the discussion and conclusions. We have included appendices to discuss (a) Backward Vs Forward gain for three-level emitter based gain medium with initial population inversion, (b) $\Lambda$-configuration with bi-directional incoherent pump between the lower to (dipole forbidden transition), (c) $\Xi$-configuration with uni-directional incoherent pump from lower level $|b\rangle$ to uppermost level $|c\rangle$, (d) brief review of vacuum-induced coherence between two-levels and (e) brief discussion of density matrix vs rate equations for two-level atom excited by an external coherent source. 
\section{Two-level quantum emitter based gain medium}
\begin{figure}[t]
  \includegraphics[height=4.7cm,width=0.28\textwidth,angle=0]{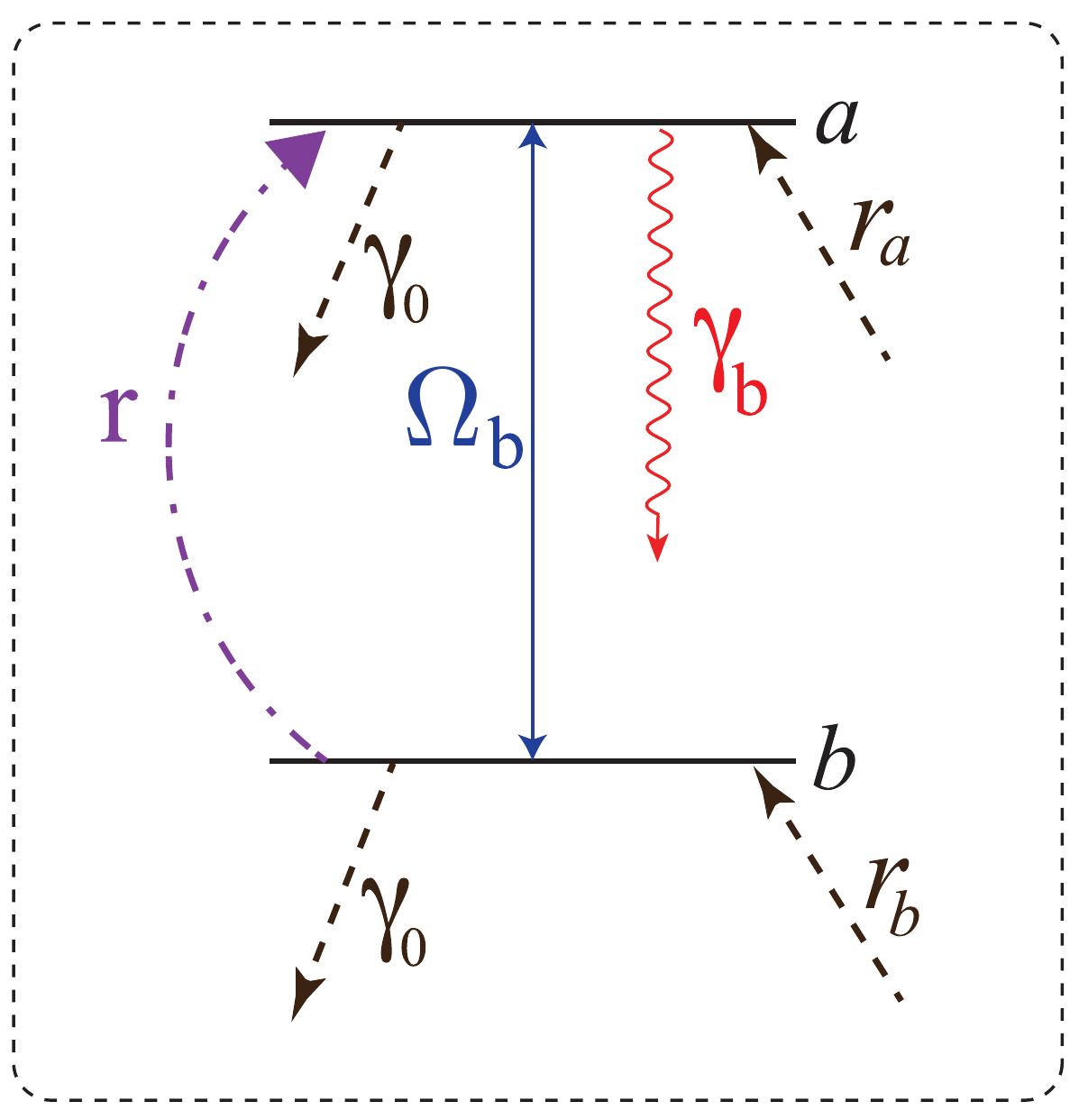}
  \caption{Two-level quantum emitter interacting with a coherent field. Field $\Omega_{b}$ couples the level  $|a\rangle$ and  $|b\rangle$. Level  $|a\rangle$ decays to  $|b\rangle$ with a rate $\gamma_{b}$ due to spontaneous emission while there is an incoherent unidirectional pumping $r$ from  $|b\rangle$ to  $|a\rangle$. $r_{a}$ and $r_{b}$ are the pumping rate into the levels $|a\rangle$ and $|b\rangle$ respectively. $\gamma_{0}$ is the decay rate out of the levels $|a\rangle$ and $|b\rangle$.}
    \label{Fig1}
\end{figure}

Let us consider the simplest yet important system in laser physics i.e a two-level quantum emitters (semiconductor quantum dots, atoms, molecules, rare-earth ions) interacting with a single mode radiation field of frequency $\nu$. Let $|a\rangle$ and $|b\rangle$ represent the upper and the lower levels of the emitter with energy $\hbar\omega_{a}$ and $\hbar\omega_{b}$ respectively (as shown in Fig.(\ref{Fig1})). We will study our system (emitter+field) in semi-classical approximation in which we treat the emitter as quantum mechanical and the field classically. To describe the the emitter response we will use the density matrix formalism. For the two-level medium the evolution of the density matrix elements $\varrho_{ij}$ takes the form~\cite{MOS}
\begin{equation}
\dot{\varrho}_{aa}=r_{a}-(\gamma_{b}+\gamma_{0})\varrho_{aa}+r\varrho_{bb}-i\left(\Omega^{\ast}_{b}\varrho_{ab}-\Omega_{b}\varrho_{ab}^{\ast}\right),
\end{equation}
\begin{equation}
\dot{\varrho}_{bb}=r_{b}+\gamma_{b}\varrho_{aa}-(r+\gamma_{0})\varrho_{bb}+i\left(\Omega^{\ast}_{b}\varrho_{ab}-\Omega_{b}\varrho_{ab}^{\ast}\right),
\end{equation}
\begin{equation}
\dot{\varrho}_{ab}=-\Gamma_{ab}\varrho_{ab}-i\Omega_{b}\left(\varrho_{aa}-\varrho_{bb}\right),
\end{equation}
where $\gamma_{b}$ is the spontaneous decay rate $|a\rangle\rightarrow |b\rangle$, $\Omega_{b}$ is the Rabi frequency for the radiation field. $\Gamma_{ab}=\gamma_{ab}+i\Delta_{b}$ is the total relaxation rate of the optical coherence which includes spontaneous emission, incoherent pumping, collisions, detunings etc. Also $\gamma_{ab}=\gamma_{0}+(\gamma_{b}+r)/2 +\gamma_{ab}^{p}$ where $\gamma_{ab}^{p}$ is the purely phase relaxation and $r$ is the rate of incoherent pump from $|b\rangle \rightarrow |a\rangle$.  $r_{a}$ and $r_{b}$ are the pumping rate into the levels $|a\rangle$ and $|b\rangle$ respectively. $\gamma_{0}$ is the decay rate out of the levels $|a\rangle$ and $|b\rangle$. In the weak probe field regime where the population of the levels do not depend on the field $\Omega_{b}$. In the this limit, the steady state ($\dot{\varrho}_{ij}=0$) populations $\varrho_{aa}$ and $\varrho_{bb}$ are given by
\begin{equation}
\varrho_{aa}^{(0)}=\frac{r_{a}(r+\gamma_{0})+r_{b}r}{\gamma_{0}(r+\gamma_{b}+\gamma_{0})},
\end{equation}
\begin{equation}
\varrho_{bb}^{(0)}=\frac{r_{b}(\gamma_{b}+\gamma_{0})+r_{a}\gamma_{b}}{\gamma_{0}(r+\gamma_{b}+\gamma_{0})}.
\end{equation}
Solving for $\varrho_{ab}$ we obtain,
\begin{equation}
\varrho_{ab}=-i\frac{\Omega_{b}}{\Gamma_{ab}}\left(\varrho_{aa}^{(0)}-\varrho_{bb}^{(0)}\right).
\end{equation}
The inversion defined as $W=\varrho_{aa}^{(0)}-\varrho_{bb}^{(0)}$ is not reached if $\varrho_{aa}^{(0)}<\varrho_{bb}^{(0)} $ which requires
\begin{equation}
\gamma_{0}(r_{a}-r_{b})<(\gamma_{b}-r)(r_{a}+r_{b}).
\end{equation}
When we do not consider any pumping into or out of the level, i.e $r_{a},r_{b},\gamma_{0}=0$, and the population is conserved i.e $\varrho_{aa}+\varrho_{bb}=1$, the emitter does not show population inversion in the steady-state if incoherent pump rate $r$ is less than the spontaneous decay rate $\gamma_{b}$. Using the definition of linear susceptibility we obtain 
\begin{equation}
\chi^{(1)}=-i\frac{3}{8\pi^{2}}N\lambda^{3}\gamma_{b}\frac{\left(\varrho_{aa}^{(0)}-\varrho_{bb}^{(0)}\right)}{\Gamma_{ab}}.
\end{equation}
The imaginary part of the complex susceptibility is given by
\begin{equation}\label{celeq9}
\text{Im}\chi^{(1)}=-\frac{3}{8\pi^{2}}N\lambda^{3}\gamma_{b}\gamma_{ab}\left(\frac{\varrho_{aa}^{(0)}-\varrho_{bb}^{(0)}}{\gamma_{ab}^{2}+\Delta^{2}}\right).
\end{equation}
The weak field $\Omega_{b}$ will be amplified if $\text{Im}\chi <0$. Thus from Eq.(\ref{celeq9}), we obtain the necessary condition for gain as 
\begin{equation}\label{celeq10}
\varrho_{aa}^{(0)}>\varrho_{bb}^{(0)}.
\end{equation}
Equation (\ref{celeq10}) is also known as population inversion condition for two-level system. It is worth mentioning here that Mollow gain\cite{Mollow72} (or hyper-Raman, or three-photon gain) can be obtained for two level system even in the absence of population inversion in the bare basis. In fact this process does not require any population in the upper level. Here two pump photons are absorbed while a probe photon is emitted. Mollow gain is a Raman-like process where the energy is transferred from the pump to the probe beam and thus it is different from amplification without inversion (AWI) where we are interested in extraction of energy from the medium. For phase effects in these hyper-Raman process in effective two-level system and using phase jump to control the excitation the readers are suggested the experimental\cite{JhaCEP1, JhaCEP2} and theoretical\cite{JhaPJ, JhaPQE13} papers respectively. It is worth mentioning here that exact solutions for the transient probability amplitudes $C_{a,b}(t)$ for two-level atoms interacting with ultra-short pulses can be obtained analytically\cite{JhaPRA10a,JhaPRA10b}.  
\section{Coherence-Enhanced Lasing I: Transient Regime}
\subsection{Three-level quantum emitter based gain medium}
In the previous section we briefly reviewed the population inversion condition required for lasing in two-level emitter based gain medium. If we add another level to the emitter, the physics of the light-matter interaction becomes rich and has opened the door for many counterintuitive physics with vast application. In this section we will review the concept of coherence-enhanced lasing in transient regime as discussed in Ref.\cite{PKJha12}.

\begin{figure}[t]
\centerline{\includegraphics[height=6cm,width=0.5\textwidth,angle=0]{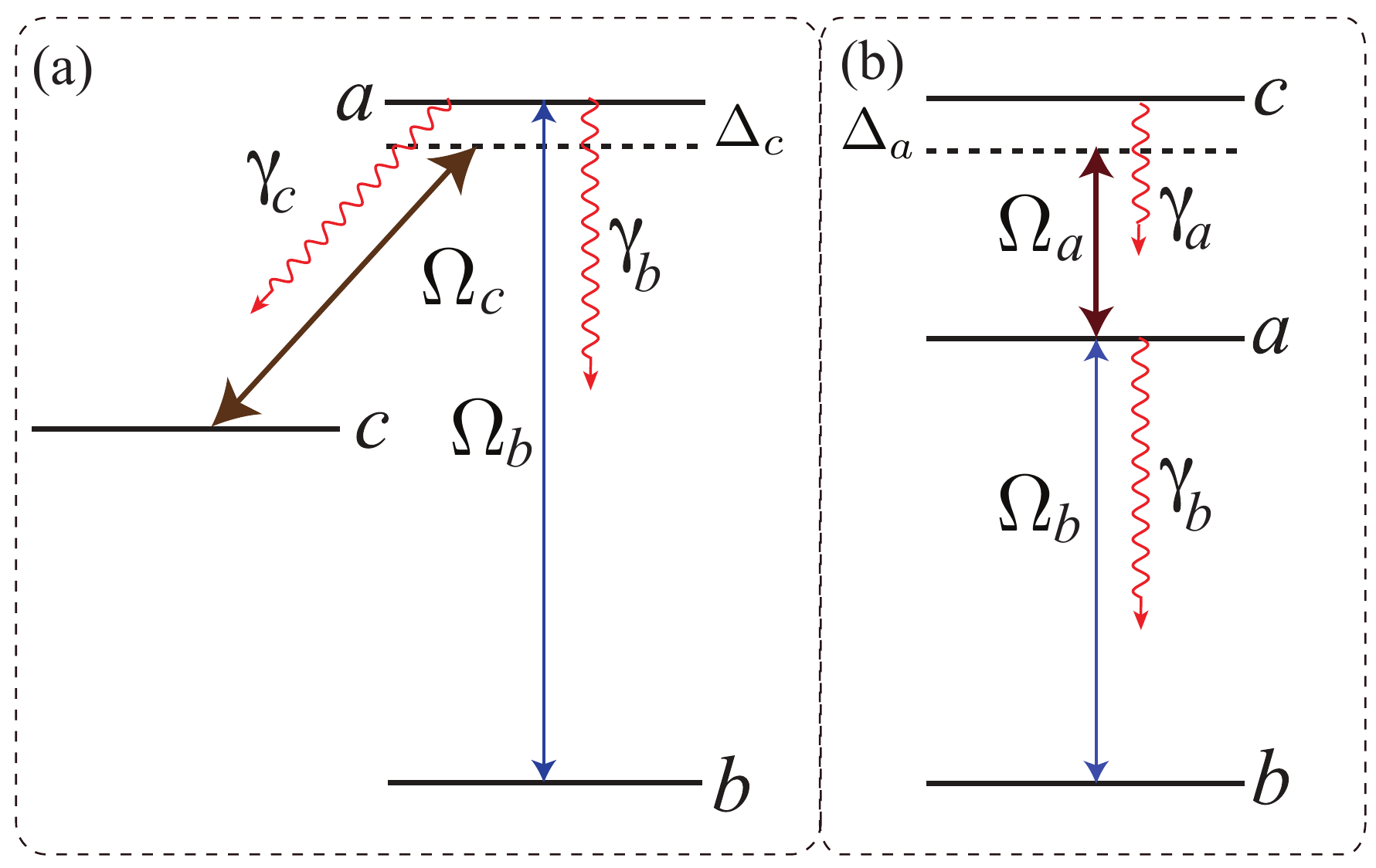}}
\caption{Three-level atomic system in (a) $\Lambda$ and (b) $\Xi$ configurations. }
\label{levels}
\end{figure}

Let us consider the three-level quantum emitter in Lambda ($\Lambda )$ configuration. Here for the sake of simplicity consider the gain medium as three-level atoms in which the transitions $|a\rangle\leftrightarrow |c\rangle$ and $|a\rangle\leftrightarrow |b\rangle$ are electric-dipole allowed but the transition $|c\rangle\leftrightarrow |b\rangle$ is electric-dipole forbidden [see Fig.~\ref{levels}] due to selection rule based on parity. We will assume that at the initial moment of time the population is distributed between levels $|a\rangle$ and $|b\rangle$ only i.e $\varrho_{aa}(0)+\varrho_{bb}(0)=1$. Any initial population in the level $|c\rangle$ is not a part of the gain medium in the absence of the drive field. Transition $|a\rangle\leftrightarrow |c\rangle$ is driven in resonance with the Rabi frequency $\Omega_{c}$. We investigate how a weak laser seed pulse at the $|a\rangle\leftrightarrow |b\rangle$ transition evolves during its propagation through the medium. The Hamiltonian in the interaction picture can be written as 
\begin{equation}\label{E1}
{\mathcal V}=\Delta_{c}|c\rangle\langle c|-[(\Omega_{b}\left |a \rangle \langle b \right |+\Omega_{c}\left |a \rangle \langle c \right |)+\text{H.c}],
\end{equation} 
 The spontaneous decay in the channels $ac$ and $ab$ are quantified by the rate $\gamma_{c}$ and $\gamma_{b}$ respectively. Incorporating these decay rates, the equation of motion for the atomic density matrix is given as
\begin{equation}\label{E2}
\begin{split}
\dot{\varrho}=-i[{\mathcal V},\varrho]+\frac{\gamma_{b}}{2}\left([\sigma_{b},\varrho\sigma_{b}^{\dagger}]+[\sigma_{b}\varrho,\sigma_{b}^{\dagger}]\right)\\+\frac{\gamma_{c}}{2}\left([\sigma_{c},\varrho\sigma_{c}^{\dagger}]+[\sigma_{c}\varrho,\sigma_{c}^{\dagger}]\right)
\end{split}
\end{equation}
where, 
\begin{equation}\label{E3}
 \sigma_{b}=\left |b \rangle \langle a \right |, \sigma_{b}^{\dagger}=\left |a \rangle \langle b \right |,  \text{and}\,  \sigma_{c}=\left |c \rangle \langle a \right |, \sigma_{c}^{\dagger}=\left |a \rangle \langle c \right |, 
\end{equation}
Here $\Omega _{b}$ and $\Omega _{c}$ are the Rabi frequencies of the probe and drive fields respectively. Evolution of the atomic density matrix $\varrho _{ij}$ is described by the set of coupled equations~\cite{MOS}
\begin{equation}
\dot{\varrho}_{ab}=-\Gamma _{ab}\varrho _{ab}-i\Omega _{b}(\varrho_{aa}-\varrho _{bb})+i\Omega _{c}\varrho _{cb},  \label{eq3}
\end{equation}
\begin{equation}
\dot{\varrho}_{cb}=-\Gamma _{cb}\varrho _{cb}+i(\Omega _{c}^{\ast }\varrho _{ab}-\Omega _{b}\varrho_{ac}^{\ast }),  \label{eq4}
\end{equation}
\begin{equation}
\dot{\varrho}_{ac}=-\Gamma _{ac}\varrho _{ac}-i\Omega _{c}(\varrho_{aa}-\varrho _{cc})+i\Omega _{b}\varrho _{cb}^{\ast },  \label{eq5}
\end{equation}
\begin{equation}
\dot{\varrho}_{bb}=\gamma _{b}\varrho _{aa}+i\left(\Omega _{b}^{\ast }\varrho _{ab}-\Omega _{b}\varrho^{\ast } _{ab}\right) ,
\label{eq6}
\end{equation}
\begin{equation}
\dot{\varrho}_{cc}=\gamma _{c}\varrho _{aa}+i(\Omega _{c}^{\ast }\varrho_{ac}-\Omega _{c}\varrho^{\ast }_{ac}),  \label{eq7}
\end{equation}
\begin{equation}
\varrho _{aa}+\varrho _{bb}+\varrho _{cc}=1,  \label{eq8}
\end{equation}
where $\Gamma _{ab}=(\gamma _{c}+\gamma _{b})/2, \Gamma _{ac}=(\gamma _{c}+\gamma _{b})/2+i\Delta_{c}, \Gamma_{cb}=-i\Delta_{c}$ are the relaxation rates of the off-diagonal elements of the atomic density matrix.
Here we assume that the spontaneous decay rate on the drive transition ($|a\rangle \rightarrow |c\rangle$) is much larger than the probe transition ($|a\rangle \rightarrow |b\rangle$). In this limit, the coherent drive will redistribute the population between the two-levels $|a\rangle$ and $|c\rangle$ and induce a non-zero coherence between them. For simplicity we will assume that  the drive field is so strong that any variation can neglected i.e $\Omega _{c}=$const (a real number), $\Omega_{b}$ is very small and the drive transition is excited resonantly $(\Delta_{a}=0)$. In this limit we can find analytical expression for the coherence $\bar{\varrho}_{ac}$ and the populations $\bar{\varrho}_{ii}$. The equation of motion is given as (putting $\gamma_{b}=0$) 
\begin{equation}
\dot{\varrho}_{cc}=\gamma _{c}\varrho _{aa}+i\Omega _{c}(\varrho_{ac}-\text{c.c}),
\end{equation}
\begin{equation}
\dot{\varrho}_{ac}=-\Gamma_{ac}\varrho_{ac}-i\Omega _{c}(\varrho _{aa}-\varrho _{cc}).
\end{equation}
\begin{equation}
\varrho _{aa}+\varrho _{cc}=\varrho _{aa}(0),
\end{equation}
The steady state solution ($\bar{\varrho}_{ij} $) can be easily obtained as\cite{PKJha12} 
\begin{equation}
\bar{\varrho} _{aa}=\frac{4\Omega _{c}^{2}}{\gamma _{c}^{2}+8\Omega _{c}^{2}}\varrho _{aa}(0), 
\end{equation}
\begin{equation}
\bar{\varrho}  _{cc}=\frac{\gamma _{c}^{2}+4\Omega _{c}^{2}}{\gamma_{c}^{2}+8\Omega _{c}^{2}}\varrho _{aa}(0),
\end{equation}
\begin{equation}
\bar{\varrho} _{ac}=\frac{2i\gamma _{c}\Omega _{c}}{\gamma _{c}^{2}+8\Omega_{c}^{2}}\varrho _{aa}(0),\label{g59}
\end{equation}
where $\varrho _{aa}(0)$ is the initial population of the level $|a\rangle$.  Evolution of the weak laser pulse $\Omega _{b}$ is described by the coupled Maxwell-Schrodinger equations 
\begin{equation}
\frac{\partial \Omega _{b}}{\partial z}+\frac{1}{c}\frac{\partial \Omega _{b}}{\partial t}=i\eta_{ab} \varrho _{ab},  \label{g60}
\end{equation}
\begin{equation}
\dot{\varrho}_{ab}=-\Gamma_{ab}\varrho _{ab}-i\Omega_{b}(\bar{\varrho} _{aa}-\bar{\varrho}_{bb})+i\Omega _{c}\varrho _{cb}  \label{g5}
\end{equation}
\begin{equation}
\dot{\varrho}_{cb}=i(\Omega _{c}\varrho _{ab}-\Omega _{b}\varrho _{ac}^{\ast }).
\label{g6}
\end{equation}
where $\eta_{ab}  =(3/8\pi )N\lambda _{ab}^{2}\gamma _{b}$ is the coupling constant, $N$ is the atomic density and $\lambda _{ab}$ is the wavelength of the $|a\rangle\leftrightarrow |b\rangle$ transition.  Here we also assumed that the drive transition has been excited resonantly $(\Delta_{c}=0)$. Next we move to find the dispersion relation by looking for solution of Eqs. (\ref{g60}-\ref{g6}) in the form
\begin{equation}
\Omega _{b}(t,z)\sim e^{i\Delta_{b} t-ik_{b}z}  \label{g63}
\end{equation}
\begin{equation}
\varrho _{ab}(t,z)\sim e^{i\Delta_{b} t-ik_{b}z}  \label{g63a}
\end{equation}
\begin{equation}
\varrho_{cb}(t,z)\sim e^{i\Delta_{b} t-ik_{b}z}  \label{g63b}
\end{equation}
which yields
\begin{equation}
\begin{split}
\left( \Delta^{2}_{b}-\Omega _{c}^{2}-\frac{i\gamma _{c}\Delta_{b} }{2}\right)&\left( ck_{b}-\Delta_{b} \right) +c\Delta_{b} \eta _{ab}(\bar{\varrho}  _{bb}-\bar{\varrho} _{aa})\\+c\eta _{ab}\Omega _{c}\bar{\varrho} _{ac}=0,
\end{split}
\label{g64}
\end{equation}
here $\Delta_{b} $ is the detuning of the laser pulse frequency from the $|a\rangle\leftrightarrow |b\rangle$ transition frequency. Now the imaginary part of $k_{b}$ determines the  gain (absorption) and takes the form
\begin{equation}
\text{Im}(k_{b})=\eta _{ab}\frac{\gamma _{c}\Delta^{2}_{b}(\bar{\varrho}  _{aa}-\bar{\varrho} _{bb})/2+\Omega _{c}\left( \Omega _{c}^{2}-\Delta^{2}_{b}\right) \text{Im}(\bar{\varrho}_{ac})}{\left( \Delta^{2}_{b}-\Omega _{c}^{2}\right) ^{2}+(\gamma_{c}\Delta_{b}/2)^{2}}.
\label{m1}
\end{equation}
In particular, for the mode resonant with the $|a\rangle\leftrightarrow |b\rangle$ transition $\Delta_{b}=0$ and we obtain
\begin{equation}
G=\text{Im}(k_{b})=\frac{\eta _{ab}}{\Omega _{c}}\text{Im}(\bar{\varrho}_{ac}).
\label{m2}
\end{equation}
When we look at the expression for $\bar{\varrho}_{ac}$, according to Eq. (\ref{g59}), Im$(\bar{\varrho} _{ac})>0$ which implies that we have gain in this configuration even when we have very little population in the upper level $|a\rangle$. Next we will study the regime in which the spontaneous decay rate on the drive transition ($|a\rangle \rightarrow |c\rangle$) is much smaller than the probe transition ($|a\rangle \rightarrow |b\rangle$). In this limit we have to consider the transient behavior of the populations and coherences to study the gain. 

Before we discuss this scenario let us briefly review the essence of gain in transient regime. The equations of motion for the density matrix element $\varrho_{ab}$ is given by Eq.(\ref{eq3}), which in the steady-state gives
\begin{equation}\label{EQ5}
\Gamma_{ab}\Im[\bar{\varrho}_{ab}]=-i\Omega_{b}(\bar{\varrho}_{aa}-\bar{\varrho}_{bb})+i\Omega_{c}\bar{\varrho}_{cb}
\end{equation}
where $\bar{\varrho}_{ij}$ is the steady-state value of the density matrix elements $\varrho_{ij}$. Now let us draw some general conclusions from the density matrix equations. From Eq.(\ref{EQ5}) it appears that, for two-level model i.e $\Omega_{c}=0$, amplification condition $\left(\Im[\bar{\varrho}_{ab}]<0\right)$ requires population inversion $\bar{\varrho}_{aa}> \bar{\varrho}_{bb}$. On the other hand for three-level system for sufficiently negative $\Omega_{c}\,\bar{\varrho}_{cb}$, the necessary condition for the probe transition $|a\rangle \leftrightarrow |b\rangle$ to exhibit amplification can be satisfied even without population inversion.
\noindent If we look at the evolution equation for the population in level $|b\rangle$,
\begin{equation}\label{EQ6}
2\Omega_{b}\Im[\varrho_{ab}]=\gamma_{b}\varrho_{aa}-\frac{\partial \varrho_{bb}}{\partial t}
\end{equation}
Amplification condition $(\Im[\varrho_{ab}] <0)$, reduces Eq.(\ref{EQ6}) to
\begin{equation}\label{EQ37}
\gamma_{b}\varrho_{aa}<\frac{\partial \varrho_{bb}}{\partial t}
\end{equation}
From Eq.(\ref{EQ37}) we conclude that in transient regime, the probe transition exhibits amplification when the transient growth of the population in level $|b\rangle$ exceeds due to the incoherent radiative decay $|a\rangle \rightarrow |b\rangle$.  Thus a net amplification can be realized in during the time of interaction $t$ iff \cite{TLWI1}
\begin{equation}
\varrho_{bb}(t)>\varrho_{bb}(0)-\gamma_{b}\int_{0}^{t}\varrho_{aa}(t')dt'
\end{equation}
In steady-state Eq.(\ref{EQ37}) can never be satisfied but it can be realized in the transient regime~\cite{TLWI1,TLWI2}. To quantify the gain(absorption) profile for the lasing transition, we analyze the parameter $G_{ab}(t)$ defined as 
\begin{equation}\label{EQ39}
G_{ab}(t)=-\frac{3}{8\pi}N\lambda^{2}_{b}\gamma_{b}\frac{\text{Im}\varrho_{ab}(t)}{\Omega_{b}}
\end{equation}
\begin{figure}[t]
\centerline{\includegraphics[height=5.5cm,width=0.48\textwidth,angle=0]{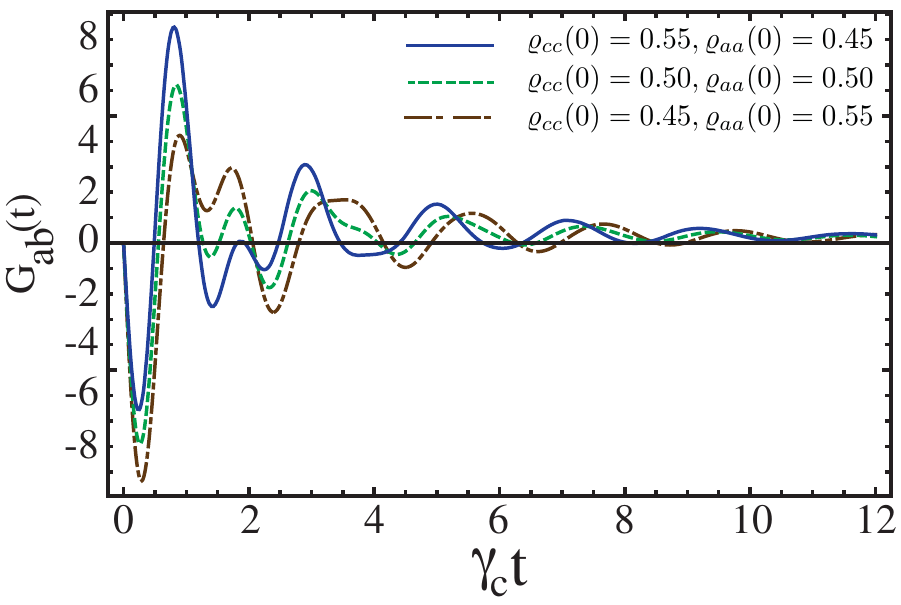}}
\caption{Plot of the gain $G_{ab}(t)$ as function of normalized time $\gamma_{c} t$. Solid line (Blue) is for initial Raman Inversion i.e $\varrho_{cc}(0) > \varrho_{bb}(0)$ while the dashed line (Green) is for zero initial Raman inversion i.e $\varrho_{bb}(0) = \varrho_{cc}(0)$ and dashed dot line (Brown) is for initial no Raman inversion i.e $\varrho_{cc}(0) < \varrho_{bb}(0)$. The common parameters taken are $\gamma_{c}=1,\gamma_{b}=0.2, \Omega_{c}=3,\Omega_{b}=0.01,\varrho_{aa}(0)=0$. For initial Raman inverted medium $\varrho_{cc}(0) =0.55, \varrho_{bb}(0)=0.45$ and $\varrho_{cc}(0) =0.45, \varrho_{bb}(0)=0.55$ for non-inverted case. We have normalized the rates and Rabi frequencies with respect to the decay rate on the drive transition $|a\rangle \rightarrow |c\rangle$. }
\label{Fig3}
\end{figure}
Here $N$ is the number density of atom (ions) and $\lambda_{b}$ is the wavelength corresponding to $|a\rangle \rightarrow |b\rangle$ transition. In general $G_{ab}(t)$ is an oscillatory function of $t$. When $G_{ab}(t) >0$, the probe pulse will enhance while for $G_{ab}(t) <0$ it will experience attenuation. In Fig. \ref{Fig3}, we have plotted $G_{ab}(t)$ for three choices of Raman inversion (positive, zero, and negative) respectively. 

In the limit $\gamma_{b} \gg \gamma_{c} $ we can find approximate analytical expression for the temporal evolution of the coherence $\varrho_{ac}$ and the populations $\varrho_{ij}$ for a weak probe field $\Omega _{b}$ and strong drive field. We obtain the solutions as
\begin{equation}  \label{eq9}
\begin{split}
\varrho _{aa} =e^{-\gamma _{b}t/2}\varrho _{aa}(0)\left\{ \sin^{2}(\Omega _{c}t)+\cos (2\Omega _{c}t)\right. \\
\left.-\frac{\gamma _{b}}{4\Omega _{c}}\sin (2\Omega_{c}t)\right\},
\end{split}
\end{equation}
\begin{equation}  \label{eq10}
\begin{split}
\varrho _{cc} =e^{-\gamma _{b}t/2}\varrho _{aa}(0)\sin^{2}(\Omega _{c}t),
\end{split}
\end{equation}
\begin{equation}  \label{eq11}
\begin{split}
\varrho _{ac}=ie^{-\gamma _{b}t/2}\varrho _{aa}(0)\sin (\Omega _{c}t)\left\{\frac{\gamma _{b}}{4\Omega _{c}}\sin(\Omega _{c}t)\right.\\
\left.-\cos (\Omega_{c}t)\right\}.
\end{split}
\end{equation}
Defining the population inversion $W(t)=\varrho_{aa}(t)-\varrho _{bb}(t)$ we obtain the expression
\begin{equation}  \label{eq12}
\begin{split}
W(t)=\frac{\varrho _{aa}(0)}{2}e^{-\gamma _{b}t/2}&\left[ 3+\cos(2\Omega _{c}t)-\right.\\
&\left.\frac{\gamma _{b}}{\Omega _{c}}\sin (2\Omega _{c}t)\right]-1.
\end{split}
\end{equation}
\begin{figure}[t]
\centerline{\includegraphics[height=5cm,width=0.49\textwidth,angle=0]{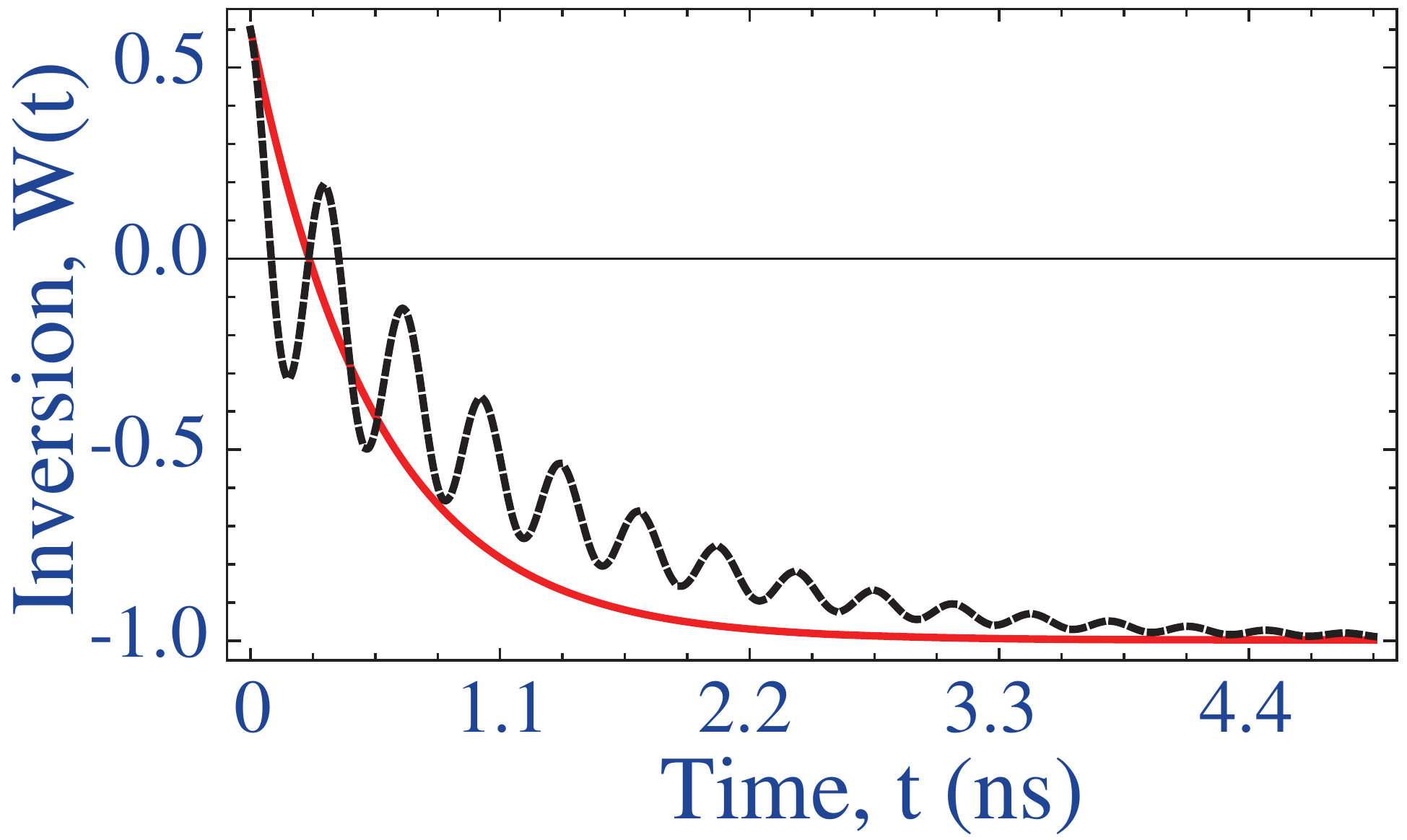}}
\caption{Inversion $W(t)$ in the probe transition ($a\leftrightarrow b$) vs time $t$. Dashed black curve shows the result for $\Omega _{c}=5\protect\gamma _{b}$ while solid red curve is obtained with no drive. For numerical simulation we used $\protect\gamma _{c}=1.1\times 10^{-3}\protect\gamma _{b}$ and the initial condition $\protect\varrho _{aa}(0)=0.8,$ $\protect\varrho _{bb}(0)=0.2$, $\protect\varrho _{cc}(0)=\protect\varrho_{ac}(0)=\protect\varrho _{ab}(0)=0$.}
\label{CETLFig5}
\end{figure}
In Fig.~\ref{CETLFig5} we plot the population difference $W(t)$ as a function of time for initial conditions $\varrho _{aa}(0)=0.8,$ $\varrho _{bb}(0)=0.2$ and $\varrho_{cc}(0)=0$. Solid line is obtained for $\Omega _{c}=5\gamma_{b}$ while for dashed line $\Omega _{c}=0$. Driving the $|a\rangle\leftrightarrow |c\rangle$ transition yields oscillations in the population difference between $|a\rangle$ and $|b\rangle$ levels. The oscillations are due to the population oscillating  between the levels $|a\rangle$ and $|c\rangle$ in the presence of the coherent drive.
\subsection{Helium as the gain medium}
\begin{figure}[t]
\centerline{\includegraphics[height=5cm,width=0.5\textwidth,angle=0]{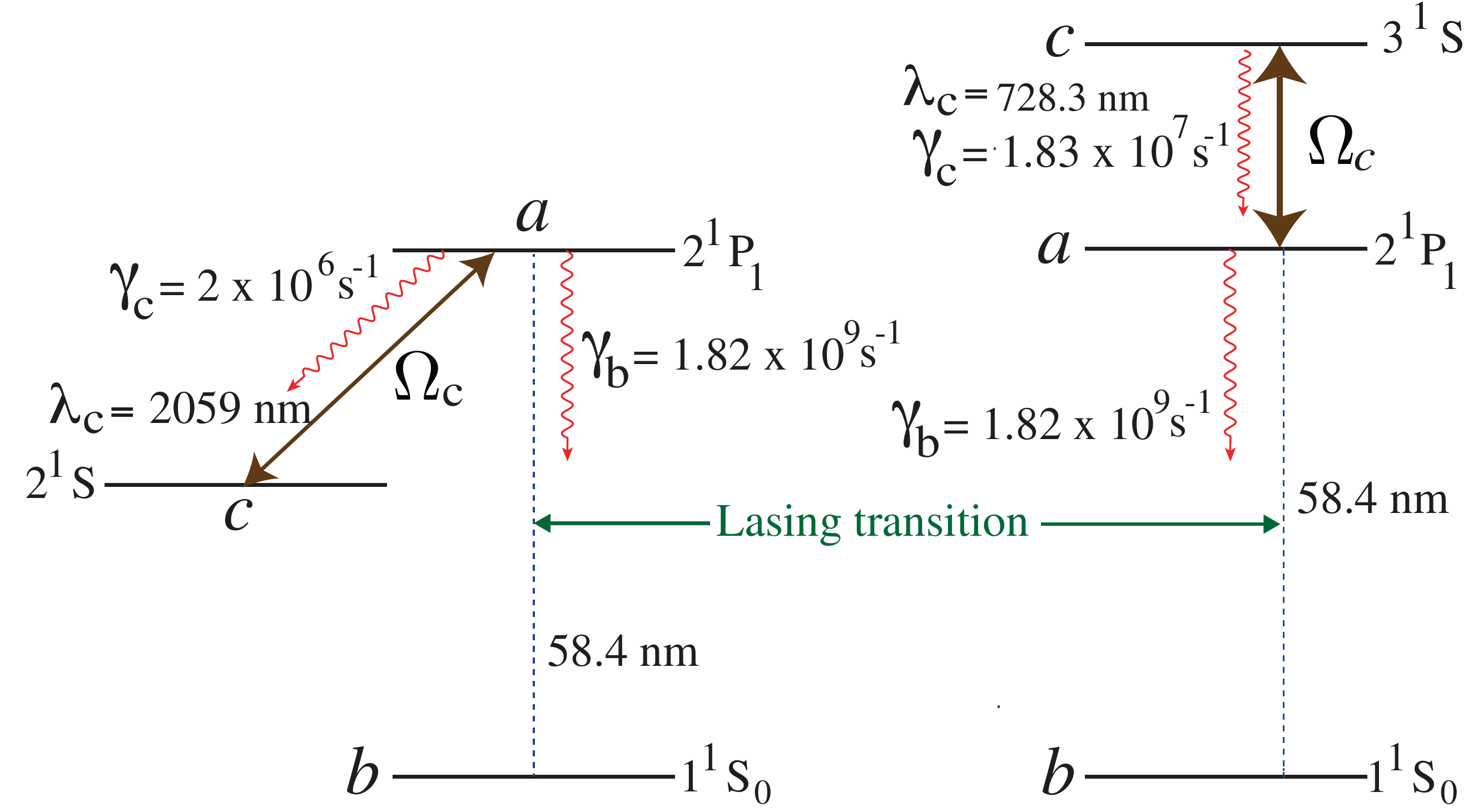}}
\caption{Energy level diagram of He atom in $\Lambda$ (left) and $\Xi$ configurations (right) respectively.  Lasing at 58.4nm can be enhanced using a coherent drive at 2059nm($\Lambda$-configuration) or 728.3nm($\Xi$-configuration)\cite{PKJha12}}
\label{CETLFig2}
\end{figure}

\begin{figure}[t]
\centerline{\includegraphics[height=5cm,width=0.5\textwidth,angle=0]{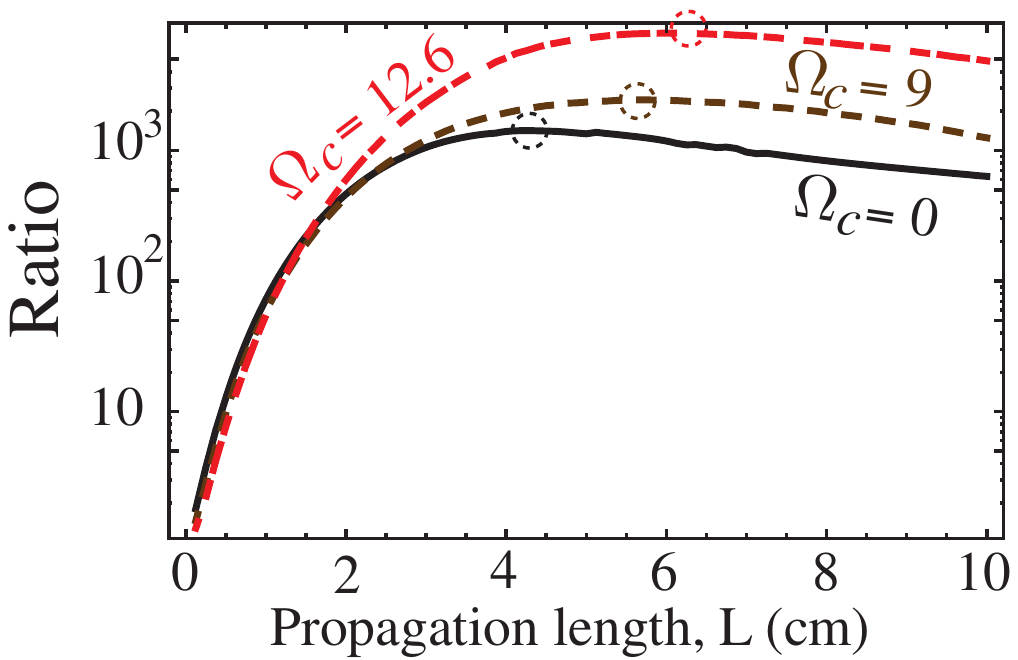}}
\caption{Ratio of the output energy to the input energy of the probe laser pulse as a function of sample length $L$ with three choices of drive Rabi frequency $\Omega_{c}=0, 9$ and $12.6\times 10^{9}\text{s}^{-1}$. In numerical simulations we take $\protect\gamma _{c}=1.1\times 10^{-3}\protect\gamma _{b}$, $\protect\eta /\protect\gamma _{b}=40.75$ cm$^{-1}$ and assume Gaussian initial probe pulse shape given by Eq. (\protect\ref{eq44}). Initial populations are $\protect\varrho _{aa}(0)=0.8,$ $\protect\varrho_{bb}(0)=0.2$ and $\protect\varrho _{cc}(0)=0 $, while initial coherences are equal to zero.}
\label{CETLFig6}
\end{figure}

\begin{figure}[b]
\centerline{\includegraphics[height=5cm,width=0.49\textwidth,angle=0]{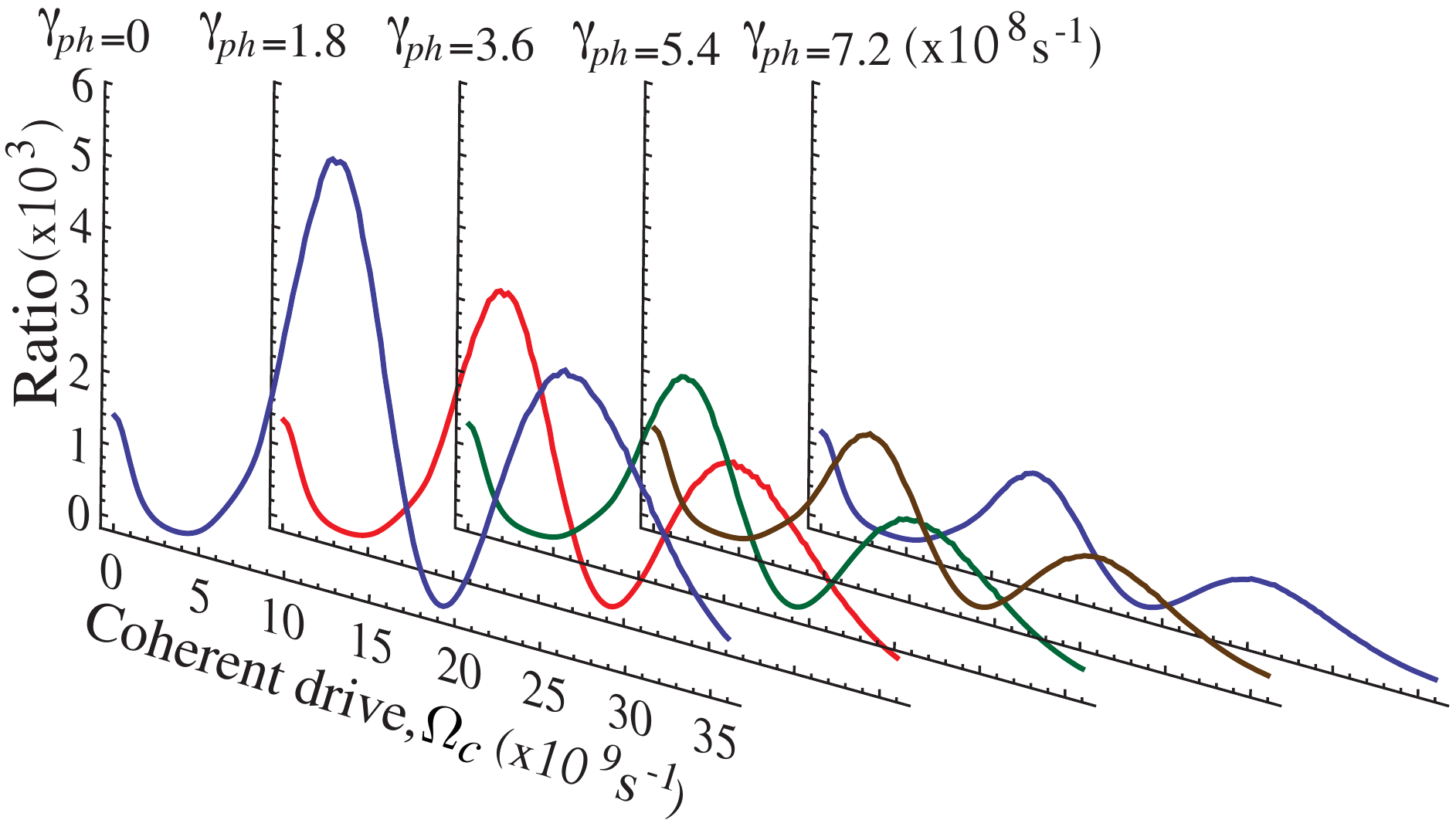}}
\caption{(a) Ratio of the output energy to the input energy of the probe laser pulse as a function of the driving field Rabi frequency $\Omega _{c}$. The ratio is $\sim 1.42 \times 10^{4}$ at $\Omega_{c}=0$. In numerical simulations we take $\protect\gamma _{c}=1.1\times 10^{-3}\protect\gamma _{b}$, $\protect\eta /\protect\gamma _{b}=40.75$ cm$^{-1}$ and assume Gaussian initial probe pulse shape given by Eq. (\protect\ref{eq44}). The length of the sample is $L=4.25$ cm, while the initial populations are $\protect\varrho _{aa}(0)=0.8,$ $\protect\varrho _{bb}(0)=0.2 $, $\protect\varrho_{cc}(0)=0$ and $\protect\varrho _{ac}(0)=\protect\varrho _{ab}(0)=0$. Other curves corresponds to different choices of the dephasing rates $\gamma_{ph}=1.8, 3.6, 5.4$ and $7.2\times 10^{8}\text{s}^{-1}$.}
\label{CETLFig7}
\end{figure}

Next we solve Eqs. (\ref{eq3})-(\ref{eq8}) and (\ref{g60}) numerically and obtain evolution of the probe laser pulse $\Omega _{b}(t,z)$ when the $|a\rangle\leftrightarrow |c\rangle$ transition is driven by a constant coherent field $\Omega _{c}$. We perform simulations for the initial condition $\varrho _{aa}(0)=0.8,$ $\varrho_{bb}(0)=0.2,$ $\varrho _{cc}(0)=0$ and take $\eta /\gamma _{b}=40.75$ cm$^{-1}$ and $\gamma _{c}=1.1\times 10^{-3}\gamma _{b}$. As an example, we consider He for which states 2$^{1}$S$_{0}$ ($|c\rangle-$ level), 2$^{1}$P$_{1}$ ($|a\rangle-$level) and the ground state 1$^{1}$S$_{0}$ ($|b\rangle-$level) form $\Lambda -$scheme [see Fig.~\ref{CETLFig2}]. The model parameters are $\lambda _{ab}=58.4$ nm, $\lambda _{ac}=2059$ nm, $\gamma_{c}=2\times 10^{6}$ s$^{-1}$ and $\gamma _{b}=1.82\times 10^{9}$ s$^{-1}$. Then for ion density $N=10^{18}$ cm$^{-3}$ we obtain $\eta /\gamma_{b}=40.75$ cm$^{-1}$.
We assume that input probe laser pulse has a Gaussian shape
\begin{equation}  \label{eq44}
\Omega _{b}(t,z=0)=0.01\exp \left[ -\left( \frac{\gamma _{b}t-0.12}{0.05}\right) ^{2}\right] \gamma _{b}.
\end{equation}
During propagation of the weak laser pulse through the medium the atomic population spontaneously decays into the ground state. After a certain time the medium is no longer inverted and the laser pulse begins to attenuate. Thus, there is an optimum length of the atomic sample which yields maximum enhancement of the pulse energy. For the optimum length the pulse leaves the medium at the onset of absorption. 
In Fig.~\ref{CETLFig6} we plot the ratio of the output pulse energy to the input energy as a function of the sample length for three choices of  external drive. We find that optimum length corresponding to maximum output energy without any drive is $L_{0}\simeq 4.25$ cm. This optimum value depends on the initial population inversion and the decay rates. At this optimum length the ratio of the output to the input probe field energy is $\sim 1.42\times 10^{3}$. In the presence of drive field $\Omega_{c}=12.6 \times 10^{9}\text{s}^{-1}$, this ratio at the propagation length $(L_{0})$ optimum for two level configuration increases to $5.81 \times 10^{3}$.  Hence we see that coherent drive can increase the energy of the XUV lasing field by 4-fold. To optimize this enhancement with respect to the drive field, but keep the sample length to be $L_{0}=4.25$ cm, we have plotted the ratio of the output to input energy as a function of Rabi frequency of the drive field. However its worth to mention that this length $L_{0}$ does not corresponds to the maximum gain for three-level system as clear from dotted circles which corresponds to optimum propagation length in Fig.~\ref{CETLFig7}.

In Fig.~\ref{CETLFig7} we plot the ratio of the output laser pulse energy (at $z=L_{0}$) to the input energy (at $z=0$) as a function of strength of the driving field $\Omega _{c}$. One can see that in the presence of coherent drive the output pulse energy oscillates as a function of $\Omega _{c}$ and for some optimum value of $\Omega_{c}$ we observed 4-fold enhancement with drive with respect to no drive. The coherent drive change the dynamics of the system in two ways (a) it redestributes the population between the upper two levels and (b) induce Raman coherence $\varrho_{cb}$ which is the key to any lasing without inversion(LWI) schemes. For $0<\Omega_{c}<2$, the drive effectively acts as an incoherent pump i.e depopulating the upper level $|a\rangle \rightarrow |c\rangle$ and the ratio goes down from $1.41\times 10^{3} (\Omega_{c}=0)$ to $4.84\times 10^{1} (\Omega_{c}=2)$. After $\Omega_{c}>2$ the ratio starts to increase and goes back to the value of $1.42 \times 10^{3}$ at $\Omega_{c} \simeq 4.625$. Beyond this value of $\Omega_{c}$, the ratio further increases and reach the global maxima $5.80\times 10^{3}$ at $\Omega_{c}\sim 7.25$ thus an enhancement of 4-fold with respect to no drive $\Omega_{c}=0$. For $7.25<\Omega_{c}<14.625$ the ratio shows similar behavior with the local minima of $4.19\times 10^{1} (\Omega_{c}=10.625)$ and local maxima $3.78\times 10^{3} (\Omega_{c}=14.625)$. Thus, coherent drive can increase the laser pulse output energy as compared to the pulse energy with no drive. We have also plotted the ratio when we phenomenologically add a dephasing term (due to collision, inhomogeneous broadening etc) to the coherence for different values of $\gamma_{ph}$. We see that the enhancement factor decreases with increasing dephasing rate and at $\gamma_{ph}\simeq 10^{9}\text{s}^{-1}$ we do not see any enhancement due to drive. Thus, one has to cleverly reduce the dephasing rate to observe the nice coherence enhanced lasing in XUV regime. 
\section{Coherence-Enhanced Lasing II: Steady-State Regime}
In this second part of the paper, we will study coherence-enhanced lasing in steady-state regime. We will begin with the generic three-level scheme as shown in Fig.~\ref{levels}(b) i.e emitter in the Cascade/Ladder configuration. Let us also assume an incoherent pump from the ground level $|b\rangle$ to the intermediate level $|a\rangle$.  Transition $|c\rangle\leftrightarrow |a\rangle$ is driven in off-resonance by the drive field. The Hamiltonian in the interaction picture can be written as 
\begin{equation}\label{eq45}
{\mathcal V}=\Delta_{a}|c\rangle\langle c|-[(\Omega_{b}\left |a \rangle \langle b \right |+\Omega_{c}\left |a \rangle \langle c \right |)+\text{H.c}],
\end{equation} 
 The spontaneous decay in the channels $ac$ and $ab$ are quantified by the rate $\gamma_{a}$ and $\gamma_{b}$ respectively. Incorporating these decay rates, the equation of motion for the atomic density matrix is given as
\begin{equation}\label{eq46}
\begin{split}
\dot{\varrho}=-i[{\mathcal V},\varrho]+\frac{\gamma_{b}}{2}\left([\sigma_{b},\rho\sigma_{b}^{\dagger}]+[\sigma_{b}\rho,\sigma_{b}^{\dagger}]\right)\\+\frac{\gamma_{a}}{2}\left([\sigma_{a},\rho\sigma_{a}^{\dagger}]+[\sigma_{a}\rho,\sigma_{a}^{\dagger}]\right)
\end{split}
\end{equation}
where, 
\begin{equation}\label{eq47}
 \sigma_{b}=\left |b \rangle \langle a \right |, \sigma_{b}^{\dagger}=\left |a \rangle \langle b \right |,  \text{and}\,  \sigma_{a}=\left |a \rangle \langle c \right |, \sigma_{a}^{\dagger}=\left |c \rangle \langle a \right |, 
\end{equation}
Here $\Omega _{b}$ and $\Omega _{a}$ are the Rabi frequencies of the probe and drive fields respectively. Evolution of the atomic density matrix $\varrho _{ij}$ is described by the set of coupled equations~\cite{MOS}
\begin{equation}
\dot{\varrho}_{ab}=-\Gamma _{ab}\varrho _{ab}-i\Omega _{b}(\varrho_{aa}-\varrho _{bb})+i\Omega^{\ast} _{a}\varrho _{cb},  \label{eq48}
\end{equation}
\begin{equation}
\dot{\varrho}_{cb}=-\Gamma _{cb}\varrho _{cb}+i\Omega_{a}\varrho_{ab}-i\Omega_{b}\varrho_{ca},  \label{eq49}
\end{equation}
\begin{equation}
\dot{\varrho}_{ca}=-\Gamma _{ca}\varrho _{ca}-i\Omega _{a}(\varrho_{cc}-\varrho _{aa})-i\Omega^{\ast}_{b}\varrho _{cb},  \label{eq50}
\end{equation}
\begin{equation}
\dot{\varrho}_{bb}=\gamma _{b}\varrho _{aa}+i\left(\Omega _{b}^{\ast }\varrho _{ab}-\Omega _{b}\varrho^{\ast}_{ab}\right) ,\label{eq51}
\end{equation}
\begin{equation}
\dot{\varrho}_{cc}=-\gamma _{a}\varrho _{cc}-i(\Omega _{a}^{\ast }\varrho_{ca}-\Omega _{a}\varrho^{\ast}_{ca}),  \label{eq52}
\end{equation}
\begin{equation}
\varrho _{aa}+\varrho _{bb}+\varrho _{cc}=1,  \label{eq53}
\end{equation}
where $\Gamma _{ab}=(\gamma _{c}+\gamma _{b})/2, \Gamma _{ac}=(\gamma _{c}+\gamma _{b})/2+i\Delta_{a}, \Gamma_{cb}=\gamma_{a}/2+i\Delta_{a}$ are the relaxation rates of the off-diagonal elements of the atomic density matrix. \noindent The steady state solutions of the off-diagonal elements of the density matrix is calculated from Eqs. (\ref{eq48}-\ref{eq50}) by setting the time derivatives terms to zero. Solving we get,
\begin{equation}
\bar{\varrho}_{ab}=-i\Omega_{b}\left\{\frac{\bar{n}_{ab}(|\Omega_{b}|^2+\Gamma_{cb}\Gamma_{ca})+\bar{n}_{ca}|\Omega_{a}|^{2}}{|\Omega_{b}|^2\Gamma_{ab}+(\Gamma_{cb}\Gamma_{ab}+|\Omega_{a}|^2)\Gamma_{ca}}\right\},\label{eq54}
\end{equation}
\begin{equation}
\bar{\varrho}_{cb}=\Omega_{b}\Omega_{a}\left\{\frac{\bar{n}_{ab}\Gamma_{ca}-\bar{n}_{ca}\Gamma_{ab}}{|\Omega_{b}|^2\Gamma_{ab}+(\Gamma_{cb}\Gamma_{ab}+|\Omega_{a}|^2)\Gamma_{ca}}\right\},\label{eq55}
\end{equation}
\begin{equation}
\bar{\varrho}_{ca}=-i\Omega_{a}\left\{\frac{\bar{n}_{ab}|\Omega_{b}|^2+\bar{n}_{ca}(\Gamma_{cb}\Gamma_{ab}+|\Omega|^2)}{|\Omega_{b}|^2\Gamma_{ab}+(\Gamma_{cb}\Gamma_{ab}+|\Omega_{a}|^2)\Gamma_{ca}}\right\}\label{eq56}
\end{equation}
\begin{figure}[b]
\centerline{\includegraphics[height=9cm,width=0.5\textwidth,angle=0]{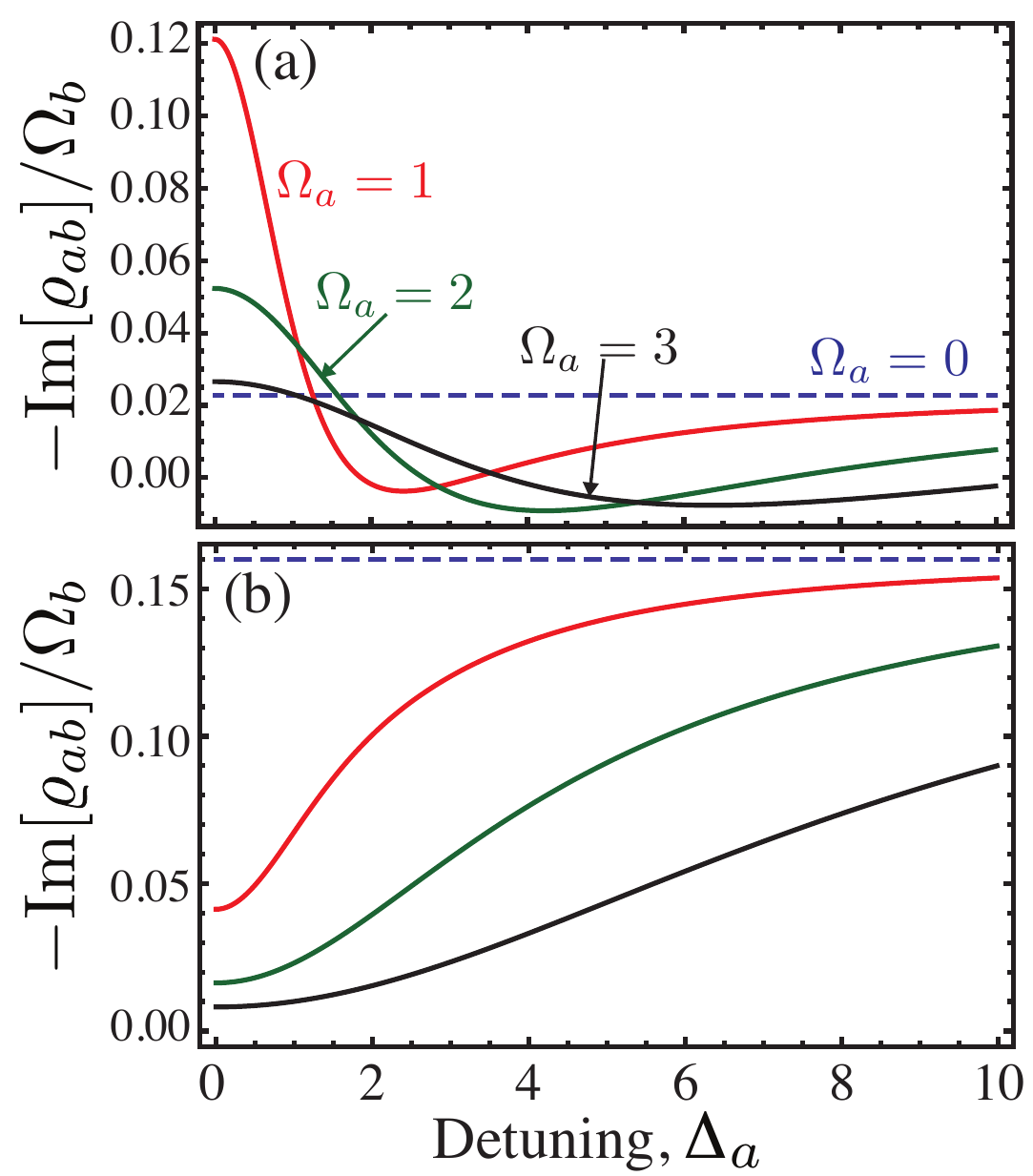}}
\caption{Plot of -\text{Im}[$\varrho_{ab}]/\Omega_{b}$ as a function of the detuning $\Delta_{a}$ for different values of the drive Rabi frequency $\Omega_{a}$. In (a) we have we assume incoherent pump from the lower level $|b\rangle \rightarrow |c\rangle$ while in (b) we considered the incoherent pump from the lower level $|b\rangle \rightarrow |a\rangle$. The dashed line corresponds to no drive $\Omega_{a}=0$.}
\label{CETLFig12}
\end{figure}
Using the condition $\Omega_{b}<<\Omega_{a}$, further reduces Eqs.(\ref{eq54}-\ref{eq56}) to,
\begin{equation}
\varrho^{(1)}_{ab}=-i\Omega_{b}\left\{\frac{n^{(0)}_{ab}\Gamma_{cb}\Gamma_{ca}+n^{(0)}_{ca}|\Omega_{a}|^{2}}{(\Gamma_{cb}\Gamma_{ab}+|\Omega_{a}|^2)\Gamma_{ca}}\right\},
\end{equation}
\begin{equation}
\varrho^{(1)}_{cb}=\Omega_{b}\Omega_{a}\left\{\frac{n^{(0)}_{ab}\Gamma_{ca}+n^{(0)}_{ca}\Gamma_{ab}}{(\Gamma_{cb}\Gamma_{ab}+|\Omega_{a}|^2)\Gamma_{ca}}\right\},
\end{equation}
\begin{equation}
\sigma^{(1)}_{ca}=-i\Omega_{a}\left\{\frac{n^{(0)}_{ca}}{\Gamma_{ca}}\right\}
\end{equation}
where $n^{(0)}_{ij}=\varrho^{(0)}_{ii}-\varrho^{(0)}_{jj}$. Here, $\varrho^{(0)}_{ii}$ is the population of the level $|i\rangle$ where we keep the probe field $\Omega_{b}$ to its lowest order while we keep all the terms for the drive field $\Omega_{a}$.  Here we have considered to scenario of unidirectional pump from (a) $|b\rangle \rightarrow |c\rangle$ and (b) $|b\rangle \rightarrow |a\rangle$. 

In Fig.(\ref{CETLFig12}) we plot the imaginary part of $\varrho_{ab}$ for the two scenarios mentioned above. When we incoherently pump to the upper level $|c\rangle$, we observed gain enhancement when we apply moderate but not strong drive. For eg when $\Omega_{a}=1$, and $\Delta_{a}=0$, the gain $\propto$ -Im$\varrho_{ab}/\Omega_{b}$ is enhanced by 6-fold. By introducing detuning on the drive transition this enhancement decreases.  When $15/8<\Delta_{a}<33/10$ the gain switches its sign and we observe loss. Thus the gain medium becomes absorptive for appropriate range of detuning $\Delta_{a}$ and can be tune back to show gain beyond this range. For $\Delta_{a}> 33/10$, transition $|a\rangle\leftrightarrow |b\rangle$ shows gain and reaches the value of gain with no drive at large detuning. Similar behavior (qualitatively) is also seen at other drive Rabi-frequency. On the other hand when we apply incoherent pump to the level $|a\rangle$ and the drive field is detuned, the gain with $\Omega_{a}\ne 0$ is lower than no drive as shown in Fig.\ref{CETLFig12}(b). When the detuning is further increased the gain reaches asymptotically to no drive gain. The dashed line in Figs. \ref{CETLFig12} corresponds to gain without drive, which should be immune to detuning as shown by constant value while varying $\Delta_{a}$. 
\subsection{Rubidium Laser}
\begin{figure}[t]
\centerline{\includegraphics[height=5cm,width=0.5\textwidth,angle=0]{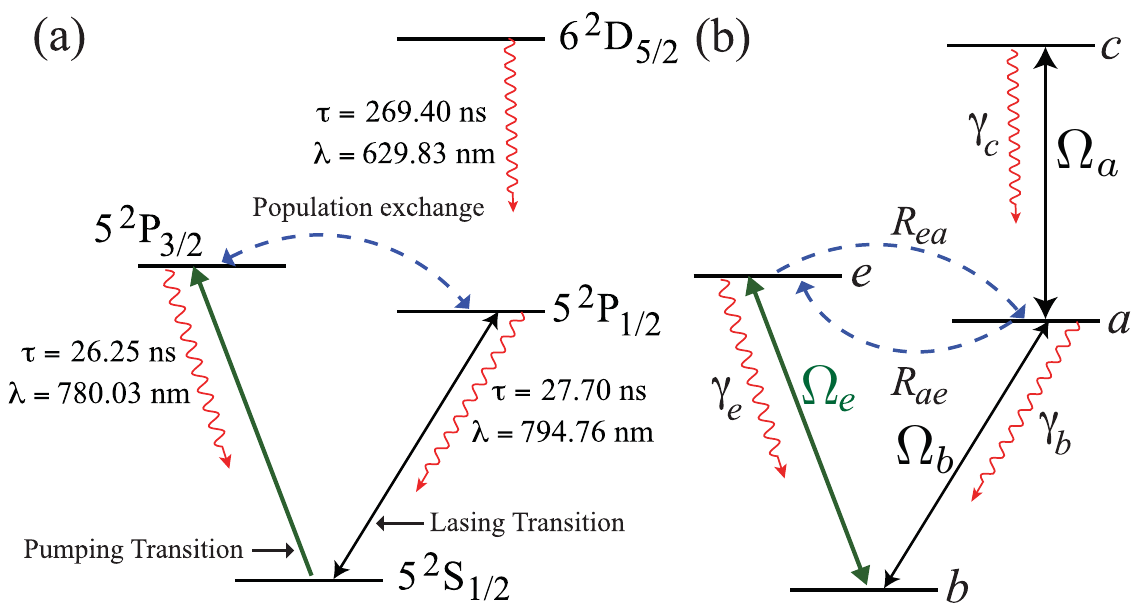}}
\caption{(a) Energy level diagram of atomic Rubidium. Here we have the dipole allowed transitions $5^{2}S_{1/2}\leftrightarrow5^{2}P_{1/2}~(D_{1})$line and $5^{2}S_{1/2}\leftrightarrow5^{2}P_{3/2}$\quad$(D_{2})$line. The population between the levels $P_{1/2}$ and $P_{3/2}$ are exchanged due to collisions by buffering the alkali vapor with other gasses like helium, ethane etc. (b) Four-level model for coherence enhanced Rubidium laser. Here couple the drive transition with a coherent field of Rabi frequency $\Omega_{a}$. The Rabi-frequency of the pump and lasing fields are $\Omega_{e}$ and $\Omega_{b}$ respectively.}
\label{Fig9}
\end{figure}

The level structure for the Rubidium laser (D$_{1}$line) is shown in Fig.~\ref{Fig9}. In the last section we discussed coherence enhanced lasing in Helium in the transient regime, here we will show that in the presence of the drive field $\Omega_{a}$ gain on the lasing transition (D$_1$line) can be enhanced substantially even in the presence of a strong dephasing rate $\gamma_{ph}$. Detailed analysis of Rubidium laser, and the conditions under which lasing action can be achieved on (D$_1$) transition ha been discussed in\cite{Payne03}.

One important condition to achieve population inversion on the lasing transition is that the rate of population exchange between $5^{2}P_{3/2}$ and $5^{2}P_{1/2}$ should be much faster than the rate of spontaneous decay from the level $5P\rightarrow 5S$. The transition $5^{2}P_{3/2} \rightarrow 5^{2}P_{1/2}$ is electric dipole forbidden, hence the population exchange is achieved using collisions with buffer gas. Here in this section we have assumed He as the buffer gas. The excitation transfer cross-section for Rb induced by collisions with rare gas atoms and alkali metals can be found in\cite{Matthew}. 

\begin{figure}[t]
\centerline{\includegraphics[height=5cm,width=0.48\textwidth,angle=0]{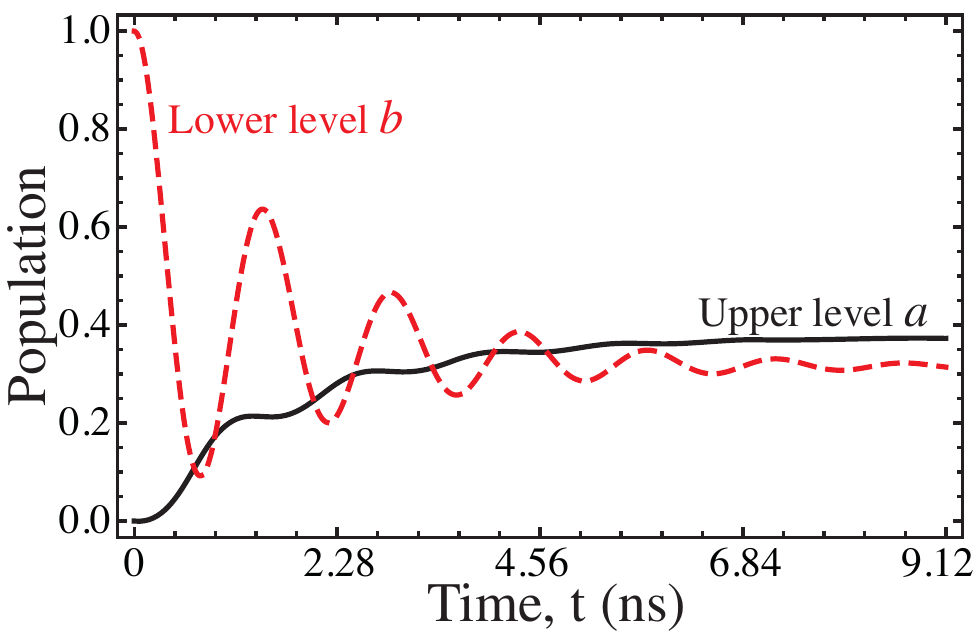}}
\caption{Temporal evolution of the populations of the level $|a\rangle$ (solid black line) and lower level $|b\rangle$ (dashed red) of the lasing transition in the absence of the drive field $\Omega_{a}$. For numerical simulation we used $\tilde{R}_{ea}=1, \tilde{R}_{ae}=0.74, \tilde{\gamma}_{b}=0.085, \tilde{\gamma}_{e}=0.087, \tilde{\gamma}_{a}=0.0085, \tilde{\Omega}_{b}=0.001, \tilde{\Omega}_{e}=5.$ Here we have normalized the decay rates and the Rabi frequencies to the population transfer rate $R_{ea}$ which is an order of magnitude higher than the spontaneous decay rates.}
\label{popu}
\end{figure}

We have modeled our gain medium as a four-level system as shown in Fig.~\ref{Fig9}(b). We excite the transition $|e\rangle \leftrightarrow |b\rangle$ by a pump laser $\Omega_{e}$ and the population between the dipole forbidden transition is exchanged via collisions with the buffer gas. In this paper we have considered ethane as the buffer gas for which we evaluated the collision rate as $R_{ae}=4.383 \times 10^{8}\text{s}^{-1}$ and  $R_{ea}=3.245 \times 10^{8}\text{s}^{-1}$. After $\sim 10$ns the system reaches steady-state with population inversion on the lasing transition as shown in Fig.(\ref{popu}). The steady-state population inversion is $\bar{n}_{ab} \simeq 0.06$ but the overall population inversion i.e $\bar{\varrho}_{aa}+\bar{\varrho}_{ee}-\bar{\varrho}_{bb} \simeq 0.369$.
\subsection{Steady State Gain}
The equation of motion for the density matrix elements $\varrho_{ij}$ are given as,
\begin{equation}\label{CETLeq50}
\dot{\varrho}_{ab}=-\Gamma_{ab}\varrho_{ab}-i\Omega_{b}(\varrho_{aa}-\varrho_{bb})+i\Omega^{\ast}_{a}\varrho_{cb}-i\Omega_{e}\varrho^{\ast}_{ea},
\end{equation}
\begin{equation}\label{CETLeq51}
\dot{\varrho}_{ca}=-\Gamma_{ca}\varrho_{ca}-i\Omega_{a}(\varrho_{cc}-\varrho_{aa})-i\Omega^{\ast}_{b}\varrho_{cb},
\end{equation}
\begin{equation}\label{CETLeq52}
\dot{\varrho}_{eb}=-\Gamma_{eb}\varrho_{eb}-i\Omega_{e}(\varrho_{ee}-\varrho_{bb})-i\Omega_{b}\varrho_{ea},
\end{equation}
\begin{equation}\label{CETLeq53}
\dot{\varrho}_{cb}=-\Gamma_{cb}\varrho_{cb}+i\Omega_{a}\varrho_{ab}-i\Omega_{b}\varrho_{ca}-i\Omega_{e}\varrho_{ce},
\end{equation}
\begin{equation}\label{CETLeq54}
\dot{\varrho}_{ce}=-\Gamma_{ce}\varrho_{ce}+i\Omega_{a}\varrho^{\ast}_{ea}-i\Omega^{\ast}_{e}\varrho_{cb},
\end{equation}
\begin{equation}\label{CETLeq55}
\dot{\varrho}_{ea}=-\Gamma_{ea}\varrho_{ea}+i\Omega_{e}\varrho^{\ast}_{ab}-i\Omega^{\ast}_{b}\varrho_{eb}-i\Omega_{a}\varrho^{\ast}_{ce}.
\end{equation}
The population terms is given as 
\begin{equation}\label{CETLeq56}
\begin{split}
\dot{\varrho}_{aa}=&-(\gamma_{b}+R_{ae})\varrho_{aa}+R_{ea}\varrho_{ee}+\gamma_{a}\varrho_{cc}\\
&+i(\Omega^{\ast}_{a}\varrho_{ca}-\Omega_{a}\varrho^{\ast}_{ca})-i(\Omega^{\ast}_{b}\varrho_{ab}-\Omega_{b}\varrho^{\ast}_{ab}),
\end{split}
\end{equation}
\begin{equation}\label{CETLeq57}
\dot{\varrho}_{bb}=\gamma_{b}\varrho_{aa}+\gamma_{e}\varrho_{ee}+i(\Omega^{\ast}_{b}\varrho_{ab}-\Omega_{b}\varrho^{\ast}_{ab})+i(\Omega^{\ast}_{e}\varrho_{eb}-\Omega_{e}\varrho^{\ast}_{eb}),
\end{equation}
\begin{equation}\label{CETLeq58}
\dot{\varrho}_{cc}=-\gamma_{a}\varrho_{cc}-i(\Omega^{\ast}_{a}\varrho_{ca}-\Omega_{a}\varrho^{\ast}_{ca}),
\end{equation}
\begin{equation}\label{CETLeq59}
\dot{\varrho}_{ee}=-(\gamma_{e}+R_{ea})\varrho_{ee}+R_{ae}\varrho_{aa}-i(\Omega^{\ast}_{e}\varrho_{eb}-\Omega_{e}\varrho^{\ast}_{eb}),
\end{equation}
where,
\begin{equation}\label{CETLeq60}
\begin{split}
\Gamma_{ca}&=\frac{\gamma_{a}+\gamma_{b}+R_{ae}}{2}, \Gamma_{ce}=\frac{\gamma_{a}+\gamma_{e}+R_{ea}}{2}\\
\Gamma_{ea}&=\frac{\gamma_{e}+\gamma_{b}+R_{ae}+R_{ea}}{2}, \quad \Gamma_{cb}=\frac{\gamma_{a}}{2},\\
\Gamma_{eb}&=\frac{\gamma_{e}+R_{ea}}{2}, \quad \Gamma_{ab}=\frac{\gamma_{b}+R_{ea}}{2}\\
\end{split}
\end{equation}
Let us assume that all the fields are real and we keep the probe field $\Omega_{b}$ to the lowest order while we consider all orders for the drive fields $\Omega_{e}$ and $\Omega_{a}$ [see Fig.~\ref{CETLFig12}(b)]. In this limit, we obtain the coherence $\varrho^{(1)}_{ab}$ as 
\begin{equation}\label{CETLeq61}
\varrho^{(1)}_{ab}=-i\Omega_{b}\left[\frac{n^{(0)}_{ab}\mathcal{A}+n^{(0)}_{ca}\mathcal{B}-n^{(0)}_{eb}\mathcal{C}}{\mathcal{D}}\right],
\end{equation}
where, the population inversion term $n^{(0)}_{ij}=\varrho^{(0)}_{ii}-\varrho^{(0)}_{jj}$ and the constants are defined as
\begin{equation}\label{CETLeq62}
\mathcal{A}=\Gamma_{ca}\Gamma_{eb}\left[\Gamma_{cb}(\Gamma_{ce}\Gamma_{ea}+\Omega^{2}_{a})+\Gamma_{ea}\Omega^{2}_{e}\right]
\end{equation}
\begin{equation}\label{CETLeq63}
\mathcal{B}=\Gamma_{eb}\left[\Gamma_{ce}\Gamma_{ea}+\Omega^{2}_{a}-\Omega^{2}_{e}\right]\Omega^{2}_{a}
\end{equation}
\begin{equation}\label{CETLeq64}
\mathcal{C}=\Gamma_{ca}\left[\Gamma_{cb}\Gamma_{ce}+\Omega^{2}_{e}-\Omega^{2}_{a}\right]\Omega^{2}_{e}
\end{equation}
\begin{equation}\label{CETLeq65}
\begin{split}
\mathcal{D}=\Gamma_{ca}\Gamma_{eb}\left[(\Gamma_{ab}\Gamma_{cb}+\Omega^{2}_{a})(\Gamma_{ce}\Gamma_{ea}+\Omega^{2}_{a})+\right. \\
\left. (\Gamma_{cb}\Gamma_{ce}+\Gamma_{ab}\Gamma_{ea}-2\Omega^{2}_{a})\Omega^{2}_{e}+\Omega^{4}_{e}\right]
\end{split}
\end{equation}
It can be easily verified that we can obtain the known results $\varrho^{(1)}_{ab}$ for cascade and Vee scheme. The zeroth order population obtained from Eqs. (\ref{CETLeq56}-\ref{CETLeq59}) as
\begin{equation}
\varrho^{(0)}_{aa}=\frac{2R_{ea}(\Gamma_{a}\Gamma_{ca}+2\Omega^{2}_{a})\Omega^{2}_{e}}{\mathcal{M}}
\end{equation}
\begin{equation}
\varrho^{(0)}_{bb}=\frac{\beta[R_{ea}\gamma_{b}\Gamma_{eb}+(R_{ae}+\gamma_{b})(\gamma_{e}\Gamma_{eb}+2\Omega^{2}_{e})]}{\mathcal{M}}
\end{equation}
\begin{equation}
\varrho^{(0)}_{cc}=\frac{2R_{ea}\Omega^{2}_{a}\Omega^{2}_{e}}{\mathcal{M}}
\end{equation}
\begin{equation}
\varrho^{(0)}_{ee}=\frac{2(R_{ae}+\gamma_{b})\Omega^{2}_{e}}{\mathcal{M}}
\end{equation}
\begin{figure}[t]
\centerline{\includegraphics[height=5cm,width=0.5\textwidth, angle=0]{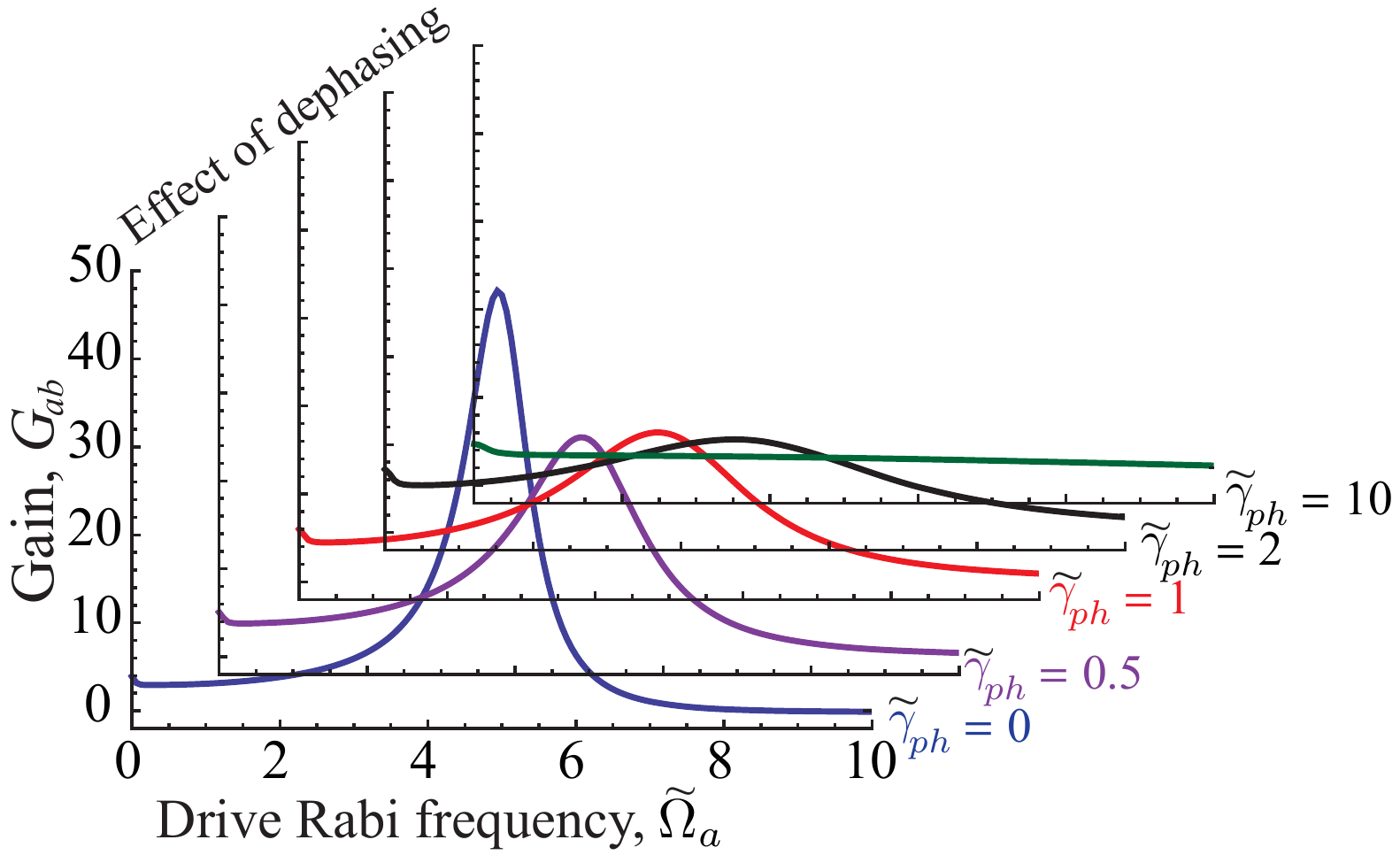}}
\caption{Plot of steady-state gain $G_{ab}$ as a function of the coherent drive $\tilde{\Omega}_{a}$ for different values of the dephasing rate $\tilde{\gamma}_{ph}$. Other curves corresponds to different choices of the dephasing rates $\tilde{\gamma}_{ph}=0.0, 0.5, 1, 2, 10$. Other parameters are same as Fig.\ref{popu}.}
\label{CETLFig14}
\end{figure}
where the constants $\beta= (\gamma_{a}\Gamma_{ca}+2\Omega^{2}_{a})$ and $ \mathcal{M}=\gamma_{a}\Gamma_{ca}\left[R_{ea}(\gamma_{b}\Gamma_{eb}+2\Omega^{2}_{e})+(R_{ae}+\gamma_{b})(\gamma_{e}\Gamma_{eb}+4\Omega^{2}_{e})\right]$. In the absence of the drive field $\tilde{\Omega}_{a}$, and strong pump field $\Omega_{e} \gg \gamma_{e},\gamma_{b},R_{ea},R_{ae}$ we obtain 
\begin{figure}[t]
\centerline{\includegraphics[height=5.0cm,width=0.5\textwidth,angle=0]{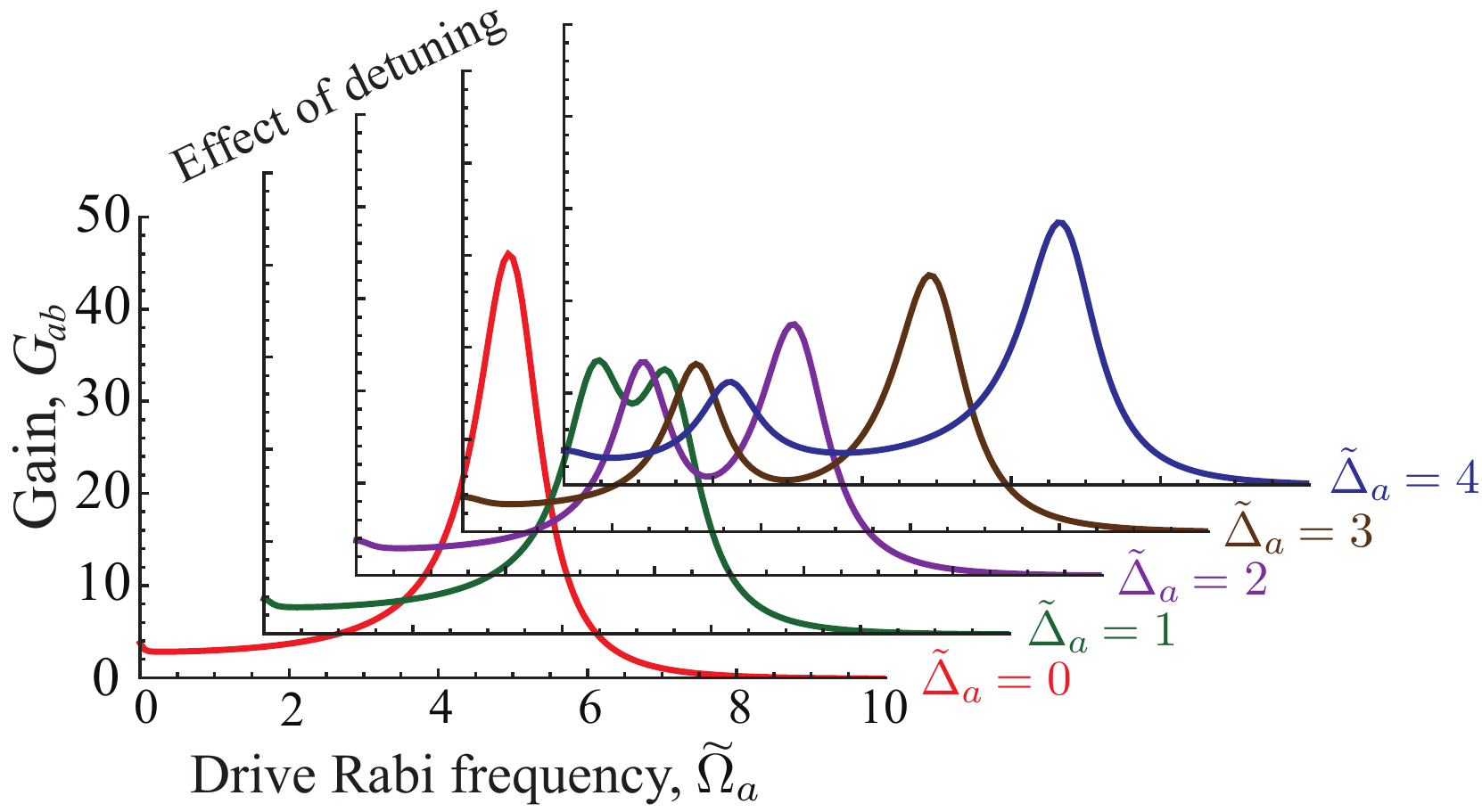}}
\caption{Plot of steady-state gain $G_{ab}$ as a function of the coherent drive $\tilde{\Omega}_{a}$ for different choices of the detuning $\tilde{\Delta}_{a}=1, 2, 3, 4$. Here $\tilde{\Delta}_{a}$ normalized to the population transfer rate $R_{ea}$. Other parameters are same as Fig. \ref{popu}.}
\label{Detuning}
\end{figure}
\begin{equation}
\varrho^{(0)}_{aa}=\frac{R_{ea}}{R_{ea}+2(R_{ae}+\gamma_{b})},
\end{equation}
\begin{equation}
\varrho^{(0)}_{bb}=\frac{R_{ae}+\gamma_{b}}{R_{ea}+2(R_{ae}+\gamma_{b})},
\end{equation}
\begin{equation}
\varrho^{(0)}_{ee}=\frac{R_{ae}+\gamma_{b}}{R_{ea}+2(R_{ae}+\gamma_{b})},
\end{equation}
The steady-state inversion ($\varrho^{(0)}_{aa}+\varrho^{(0)}_{ee}-\varrho^{(0)}_{bb}$) is given as,
\begin{equation}
\frac{R_{ae}}{R_{ea}+2(R_{ae}+\gamma_{b})}>0
\end{equation}
and also on the lasing transition $(a\leftrightarrow b)$ we obtain,
\begin{equation}
\varrho^{(0)}_{aa}-\varrho^{(0)}_{bb}= \frac{R_{ea}-R_{ae}-\gamma_{b}}{R_{ea}+2(R_{ae}+\gamma_{b})},
\end{equation}
For inversion on the lasing transition we require $R_{ea}>R_{ae}+\gamma_{b}$. This is a very crucial condition to keep in mind while selecting appropriate buffer gas for population exchange between the upper levels $|e\rangle$ and $|a\rangle$. We have also included here an appendix to review similar conditions for three-level emitter in Lambda-configuration with symmetric bi-directional pump for the convenience of the readers.

Let us now study the effect of drive field $\tilde{\Omega}_{a}$ on the steady-state gain $G_{ab}$ defined by Eq.(\ref{EQ39}). The result of the numerical simulation of Eqs.~(\ref{CETLeq50}-\ref{CETLeq59}) is shown in Fig.~\ref{CETLFig14} in which we have shown the effect of the drive field on the gain $G_{ab}$. We see that in the presence of the drive field we can enhance the gain by an order of magnitude for $\tilde{\Omega}_{a}\sim 5$. To emphasize the role of the coherence we simulated Eqs.~(\ref{CETLeq50}-\ref{CETLeq59}) in the presence of the several dephasing rates $\tilde{\gamma}_{ph}$ and the results are shown in Fig~\ref{CETLFig14}. We see that even when the dephasing rate is 20-fold higher than the spontaneous decay rate in the lasing transition, quantum coherence can still enhance the gain by 2-fold. 

In Fig.\ref{Detuning} we have plotted the steady-state gain $G_{ab}$ defined by Eq.(\ref{EQ39}) as a function of the coherent drive $\Omega_{a}$ for different choices of the detuning $\tilde{\Delta}_{a}=0,1, 2, 3, 4$. Interestingly the central peak splits into two peaks which further move apart as we increase the detuning. The first peak vanishes beyond $\tilde{\Delta}_{a}=5$. The position of the peak is also asymmetric with respect to the peak with zero detuning ($\tilde{\Delta}_{a}=0$). We also observed that within the range of parameters (drive detuning and  drive Rabi-frequency) used to simulate Fig.(12) there was no significant change in the width of the second peak.
\section{Conclusion}
In this paper, we have studied one of the manifestation of quantum coherence effects in atomic physics. We showed that quantum coherence and interference effects can be used as a tool to enhance gain in both XUV and infra-red wavelength under suitable conditions in transient and steady-state regimes respectively. We investigated the role of coherence in substantial enhancement, more than an order of magnitude, in the output energy when compared to no drive model ($\Omega_{a}=0$). 

For transient regime we selected Helium atoms (lasing at 58.4nm), initially prepared in the highly inverted state, our gain medium. We optimized the propagation length for a given population inversion by studying the energy of the seed pulse with varying propagation length. At the optimum length, the ratio of output to input energy is $1.41\times10^{3}$. At this optimum length we applied a coherent drive $(\lambda_{a} \sim 2\mu$m ) and optimized the drive Rabi-frequency to enhance the gain. For moderate drive we theoretically demonstrated a 4-fold enhancement in the ratio. For steady-state regime, we selected Rubidium atoms (lasing at 794.76nm) as our gain medium. Here we observed more than an order of magnitude enhancement in the steady-state gain even when the phase relaxation rates (dephasing) along with other decoherence rates are 20-fold faster than the spontaneous decay rate on the lasing transition. These results though seems optimistic in the XUV regime, which is the main motivation for this project, can provide a route to building powerful laser if we can cleverly handle and reduce the fast relaxation rates. \textit{ A proof of principle experiment on coherence-enhanced lasing with population inversion using favorite atoms like alkali metals (Rb, Na, K) would be one way to go.}  

With the recent proposals\cite{Dorfman13,PKJha13} on extending the coherence effects to new domain of plasmonics has definitively opened new prospects and challenges which has been unexplored till now. Moreover optical control or rather coherent control in plasmonics will add a new dimension in nanophotonics.

\section{Acknowledgement}
It is a pleasure to thank my collaborators to whom I give most of the credits. The theoretical ideas on using quantum coherence to enhance gain was developed in collaboration with Anatoly Svidzinsky and Marlan O. Scully. My sincere appreciation for all the efforts of L.V. Keldysh with whom I had countless long discussions on different aspects of laser physics while his stay at Texas A\&M University. I would like to thank  S. Suckewer, O. Kocharovskaya, M. S. Zubairy, Y. V. Rostovtsev, H. Eleuch and E. Sete for fruitful discussions. I  gratefully acknowledge financial support from the Welch Foundation Graduate Fellowship, the Herman F. Heep and Minnie Belle Heep Texas A\&M University Endowed Fund held and administered by the Texas A\&M Foundation, the Texas A\&M Research and Presentation Grant, and the American Physical Society Student Travel Grants.

\appendix
\section{Backward Vs Forward Gain}
In this paper have considered the evolution of the injected seed pulse at $z=0$ in the forward direction (with respect to the driving laser). In this section we will briefly discuss the evolution of an identical seed pulse injected at in the backward direction at the end of the sample i.e $z=L$ along with the forward seed pulse. We will consider the three-level system in Lambda configuration in the limit $\gamma_{b}\gg \gamma_{c}$ as shown in Fig.~\ref{levels} and we drive the transition $a\leftrightarrow c$ by a field $\vec{E}_{c}$.  We also excite the transition $a\leftrightarrow b$ by a weak probe field $\vec{E}_{b}$. We write the electric field as,
\begin{equation}
\vec{E}_{c}(z,t)=\frac{\epsilon^{+}_{c}}{2}\left[\mathcal{E}^{+}_{c}(z,t)e^{i\theta^{+}_{c}}+\text{c.c}\right],
\end{equation}
\begin{equation}
\vec{E}^{\pm}_{b}(z,t)=\frac{\epsilon^{\pm}_{b}}{2}\left[\mathcal{E}^{\pm}_{c}(z,t)e^{i\theta^{\pm}_{c}}+\text{c.c}\right],
\end{equation}
where 
\begin{equation}
\theta^{+}=kz-\nu t, \quad \theta^{-}=-kz-\nu t 
\end{equation}
Here (+) and (-) sign as the superscript means forward and backward direction respectively. We can write the off-diagonal term as 
\begin{equation}\label{CETLeq38}
\dot{\rho}_{ab}=-\omega_{ab}\rho_{ab}-i\vec{\wp}_{ab}\cdot\vec{E}_{b}(\rho_{aa}-\rho_{bb})+i\vec{\wp}_{ac}\cdot\vec{E}_{c}\rho_{cb}
\end{equation}
\begin{equation}\label{CETLeq39}
\dot{\rho}_{ac}=-\omega_{ac}\rho_{ac}-i\vec{\wp}_{ac}\cdot\vec{E}_{c}(\rho_{aa}-\rho_{cc})+i\vec{\wp}_{ab}\cdot\vec{E}_{b}\rho^{\ast}_{cb}
\end{equation}
\begin{equation}\label{CETLeq40}
\dot{\rho}_{cb}=-\omega_{cb}\rho_{cb}-i\vec{\wp}_{ab}\cdot\vec{E}_{b}\rho^{\ast}_{ac}+i\vec{\wp}_{ca}\cdot\vec{E}_{c}\rho_{ab}
\end{equation}
Let us make the following transformation
\begin{equation}\label{CETLeq41}
\rho_{ab}=\varrho^{+}_{ab}e^{i\theta^{+}_{1}}+\varrho^{-}_{ab}e^{i\theta^{-}_{1}},
\end{equation}
\begin{equation}\label{CETLeq42}
\rho_{ac}=\varrho^{+}_{ac}e^{i\theta^{+}_{c}},
\end{equation}
\begin{equation}\label{CETLeq43}
\rho_{cb}=\varrho^{+}_{cb}e^{i\theta^{+}_{3}}+\varrho^{-}_{cb}e^{i\theta^{-}_{3}},
\end{equation}
where
\begin{equation}\label{CETLeq44}
\theta^{\pm}_{1}=\theta^{\pm}_{b};\, \theta^{+}_{2}=\theta^{+}_{c};\,  \theta^{+}_{3}=\theta^{+}_{b} -\theta^{+}_{c};\,  \theta^{-}_{3}=\theta^{-}_{b} -\theta^{+}_{c} 
\end{equation}
Using the transformation Eqs.~(\ref{CETLeq41}-\ref{CETLeq44}) in Eqs.~(\ref{CETLeq38}-\ref{CETLeq40}) we obtain for the backward direction:
\begin{equation}\label{CETLeq45}
\dot{\varrho}^{-}_{ab}=-\Gamma_{ab}\varrho^{-}_{ab}-i\Omega^{-}_{b}(\varrho_{aa}-\varrho_{bb})+i\Omega_{c}\varrho^{-}_{cb}
\end{equation}
\begin{equation}\label{CETLeq46}
\dot{\varrho}^{-}_{cb}=-\Gamma_{cb}\varrho^{-}_{cb}-i\Omega^{-}_{b}\varrho^{+\ast}_{ac}+i\Omega^{\ast}_{c}\varrho^{-}_{ab}
\end{equation}
and for the forward direction we obtain
\begin{equation}\label{CETLeq47}
\dot{\varrho}^{+}_{ab}=-\Gamma_{ab}\varrho^{+}_{ab}-i\Omega^{+}_{b}(\varrho_{aa}-\varrho_{bb})+i\Omega_{c}\varrho^{+}_{cb}
\end{equation}
\begin{equation}\label{CETLeq48}
\dot{\varrho}^{+}_{cb}=-\Gamma_{cb}\varrho^{+}_{cb}-i\Omega^{+}_{b}\varrho^{+\ast}_{ac}+i\Omega^{\ast}_{c}\varrho^{+}_{ab}
\end{equation}
\begin{equation}\label{CETLeq49}
\dot{\varrho}^{+}_{ac}=-\Gamma_{ac}\varrho^{+}_{ac}-i\Omega_{c}(\varrho_{aa}-\varrho_{cc})+i\Omega^{+}_{b}\varrho^{+\ast}_{cb}
\end{equation}
The evolution of the population is given as 
\begin{equation}\label{CETLeq49a}
\begin{split}
\dot{\varrho}_{aa}=&-(\gamma_{b}+\gamma_{c})\varrho_{aa}-i(\Omega^{+\ast}_{b}\varrho^{+}_{ab}-\Omega^{+}_{b}\varrho^{+\ast}_{ab})-i(\Omega^{-\ast}_{b}\varrho^{-}_{ab}\\
&-\Omega^{-}_{b}\varrho^{-\ast}_{ab})-i(\Omega^{+\ast}_{c}\varrho^{+}_{ac}-\Omega^{+}_{c}\varrho^{+\ast}_{ac})
\end{split}
\end{equation}
\begin{equation}
\dot{\varrho}_{bb}=\gamma_{b}\varrho_{aa}+i(\Omega^{+\ast}_{b}\varrho^{+}_{ab}-\Omega^{+}_{b}\varrho^{+\ast}_{ab})+i(\Omega^{-\ast}_{b}\varrho^{-}_{ab}-\Omega^{-}_{b}\varrho^{-\ast}_{ab})
\end{equation}
\begin{equation}\label{CETLeq49c}
\dot{\varrho}_{cc}=\gamma_{c}\varrho_{aa}+i(\Omega^{+\ast}_{c}\varrho^{+}_{ac}-\Omega^{+}_{c}\varrho^{+\ast}_{ac})
\end{equation}
From Eqs. (\ref{CETLeq49a}-\ref{CETLeq49c}) we see that the population equations are symmetric under the transformation $ + \leftrightarrow -$, hence the evolution of the injected (identical) seed pulse at the respective ends of the sample will be the same. 

If we can selectively destroy the coherence in the backward direction ($\varrho^{-}_{ij}$), keeping the forward coherence intact, we can in principle get asymmetric results as discussed in \cite{PKJha12}. In the opposite limit, than the results reported in \cite{PKJha12}, greater energy in the backward direction will be beneficial in the sky laser physics\cite{Sky11}.
\section{Three-level lambda with bi-directional pump}
\begin{figure}[t]
\centerline{\includegraphics[height=5cm,width=0.26\textwidth,angle=0]{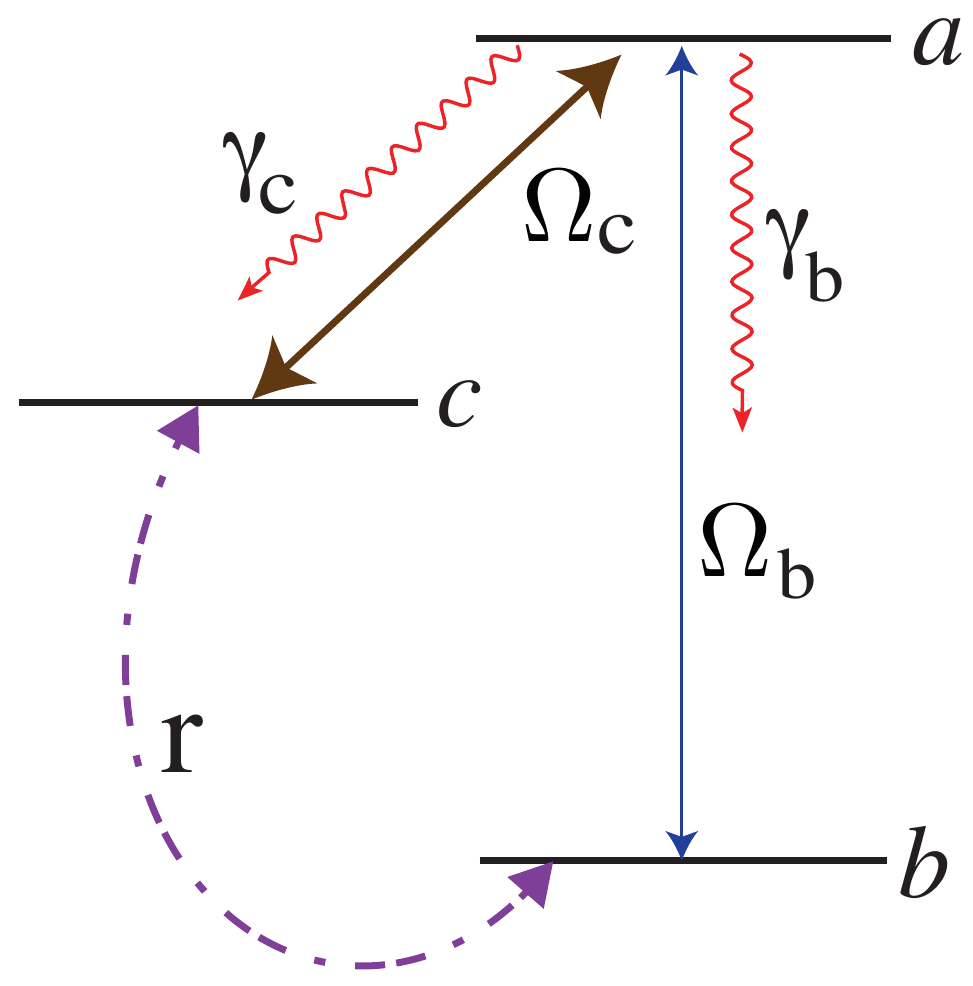}}
\caption{Three-level atomic system in $\Lambda -$configuration. Here we have included a bi-directional pump given by the rate $r$ to exchange population between the lower two levels }
\label{3ss}
\end{figure}
Let us consider three-level emitter in Lambda ($\Lambda)$ configuration as shown in Fig.13. Here we have included a bi-directional pump given by the rate $r$ to exchange population between the lower two levels. The evolution of the density matrix elements $\varrho_{ij}$ is described by the set of coupled equations~\cite{MOS}:
\begin{equation}\label{B1}
\dot{\varrho}_{ab}=-\Gamma_{ab}\varrho _{ab}+i\Omega_{b}(\varrho _{bb}-\varrho _{aa})+i\Omega _{c}\varrho _{cb},
\end{equation}
\begin{equation}\label{B2}
\dot{\varrho}_{cb}=-\Gamma_{cb}\varrho_{cb}+i(\Omega _{c}^{\ast }\varrho _{ab}-\Omega _{b}\varrho _{ac}^{\ast}),
\end{equation}
\begin{equation}\label{B3}
\dot{\varrho}_{ac}=-\Gamma_{ac}\varrho _{ac}-i\Omega_{c}(\varrho _{aa}-\varrho _{cc})+i\Omega _{b}\varrho _{cb}^{\ast },
\end{equation}
\begin{equation}\label{B4}
\dot{\varrho}_{bb}=r\varrho_{cc}-r\varrho_{bb}+\gamma_{b}\varrho_{aa}+i\left( \Omega_{b}^{\ast }\varrho _{ab}+\Omega _{b}\varrho _{ab}^{\ast }\right) ,
\end{equation}
\begin{equation}\label{B5}
\dot{\varrho}_{cc}=-r\varrho_{cc}+r\varrho_{bb}+\gamma _{c}\varrho _{aa}+i(\Omega _{c}^{\ast }\varrho _{ac}-\Omega_{c}\varrho _{ac}^{\ast }),
\end{equation}
\begin{equation}\label{B6}
\varrho _{aa}+\varrho _{bb}+\varrho _{cc}=1.
\end{equation}
where $\Gamma_{ij}=\Gamma_{ij}^{\ast}, \Gamma_{ab}=\Gamma_{ac}=(\gamma_{b}+\gamma_{c})/2$ and $\Gamma_{cb}=r$. Here $\Omega_{b}$ and $\Omega_{c}$ are the Rabi frequencies of the probe and driving field respectively. $\Gamma_{ij}$ is the relaxation rates of the off-diagonal terms $\varrho_{ij}$.  Similar to the two-level system, we obtained the steady-state (long time limit) solution for the populations $\varrho_{ll}$ in the weak probe field  and $\Omega_{c}$ is real,
\begin{equation}\label{B7}
\varrho^{(0)}_{aa}=\frac{2\Omega^{2}_{c}r}{\mathcal{Q}},
\end{equation}
\begin{equation}\label{B8}
\varrho^{(0)}_{bb}=\frac{(\gamma_{b}+\gamma_{c})r\Gamma_{ca}+2(\gamma_{b}+r)\Omega^{2}_{c}}{\mathcal{Q}},
\end{equation}
\begin{equation}\label{B9}
\varrho^{(0)}_{cc}=\frac{(\gamma_{b}+\gamma_{c})r\Gamma_{ca}+2r\Omega^{2}_{c}}{\mathcal{Q}},
\end{equation}
where $\mathcal{Q}=2r(\gamma_{b}+\gamma_{c})\Gamma_{ca}+2(\gamma_{b}+3r)\Omega^{2}_{c}$. Population inversion can be calculated from Eq.(18-20) as, 
\begin{equation}\label{B10}
\begin{split}
W=\left(\varrho^{0}_{aa} +\varrho^{0}_{cc}-\varrho^{0}_{bb}\right)
=\frac{2(r-\gamma_{b})\Omega^{2}_{c}}{\mathcal{Q}},
\end{split}
\end{equation}
From Eq.(\ref{B10}) we obtain the no population inversion\cite{Y1,Y2}  $\left(\varrho^{(0)}_{aa} +\varrho^{(0)}_{cc}-\varrho^{(0)}_{bb} <0\right)$ condition as
\begin{equation}\label{B11}
\gamma_{b}>r.
\end{equation}
We obtain $\varrho_{ab}$ as
\begin{equation}\label{B12}
\varrho_{ab}^{(1)}=-i\Omega_{b}\left\{\frac{\left[\varrho^{(0)}_{aa}- \varrho^{(0)}_{bb}\right]\Gamma_{cb}\Gamma_{ac}+\left[\varrho^{(0)}_{cc}- \varrho^{(0)}_{aa}\right]\Omega_{c}^{2}}{\left(\Gamma_{cb}\Gamma_{ab}+\Omega_{c}^2\right)\Gamma_{ac}}\right\},
\end{equation}
Substituting Eqs.(\ref{B7}-\ref{B9}) in Eq. (\ref{B12}) gives,
\begin{equation}
\varrho_{ab}^{(1)}=-i\Omega_{b}\left\{\frac{\kappa+\left[(\gamma_{b}+\gamma_{c}+2\Gamma_{cb})r-2(\gamma_{b}+r)\Gamma_{cb}\right]\Omega^{2}_{c}}{\left(\Gamma_{cb}\Gamma_{ab}+\Omega^{2}_{c}\right)\mathcal{Q}}\right\}.
\end{equation}
where $\kappa=-(\gamma_{b}+\gamma_{c})r\Gamma_{ca}$. In the strong drive field limit, to observe gain in the probe transition requires Im$\varrho_{ab} <0$ requires
\begin{equation}\label{B14}
\gamma_{c}>\gamma_{b},
\end{equation}
which means the rate of spontaneous decay in the driving transition $|a\rangle \rightarrow |c\rangle$ should be greater than probe transition $|a\rangle \rightarrow |b\rangle$. Thus to observe gain without population inversion the rates should satisfy Eqs.(\ref{B11}) and (\ref{B14}) simultaneously.
\section{Three-Level $\Xi$-configuration: Gain with uni-directional pump}
Let us now consider the cascade/ladder scheme (as shown in Fig. \ref{levels}(b)) with a uni-directional pump from the lower level $|b\rangle$ to the upper level $|c\rangle$ at a rate $\kappa$. To the zeroth order approximation in the probe field we obtained for the steady-state populations
\begin{equation}\label{C1}
\varrho^{(0)}_{aa}=\frac{\kappa(\gamma_{a}\Gamma_{ca}+2\Omega^{2}_{a})}{\mathcal{R}},
\end{equation}
\begin{equation}\label{C2}
\varrho^{(0)}_{bb}=\frac{\gamma_{b}(\gamma_{a}\Gamma_{ca}+2\Omega^{2}_{a})}{\mathcal{R}},
\end{equation}
\begin{equation}\label{C3}
\varrho^{(0)}_{cc}=\frac{\kappa(\gamma_{b}\Gamma_{ca}+2\Omega^{2}_{a})}{\mathcal{R}},
\end{equation}
where $\mathcal{R}=\kappa(\gamma_{b}\Gamma_{ca}+2\Omega^{2}_{a})+(\gamma_{b}+\kappa)(\gamma_{a}\Gamma_{ca}+2\Omega^{2}_{a})$. Here we also assumed resonant drive and probe excitation. In the strong field limit $\Omega_{a} \gg \gamma_{b},\gamma_{a},\kappa$, we obtain the non-inversion condition as 
\begin{equation}\label{C4}
\kappa < \gamma_{b}
\end{equation}
Also for gain we obtain
\begin{equation}\label{C5}
\kappa > \sqrt{\gamma_{b}\gamma_{a}}
\end{equation}
Combining Eq.(\ref{C4},\ref{C5}), the condition for gain without population inversion gives~\cite{Shuker1}
\begin{equation}\label{C6}
\sqrt{\gamma_{a}\gamma_{b}} < \kappa <\gamma_{b},\,\, \text{and}\,\,\, \gamma_{b} > \gamma_{a}
\end{equation}
We see from Eq.(\ref{C6}) that the spontaneous emission rate on the lasing and drive transition is opposite to that in Lambda-configuration.
\section{Vacuum-induced coherence}
\begin{figure}[t]
\centerline{\includegraphics[height=5cm,width=0.15\textwidth,angle=0]{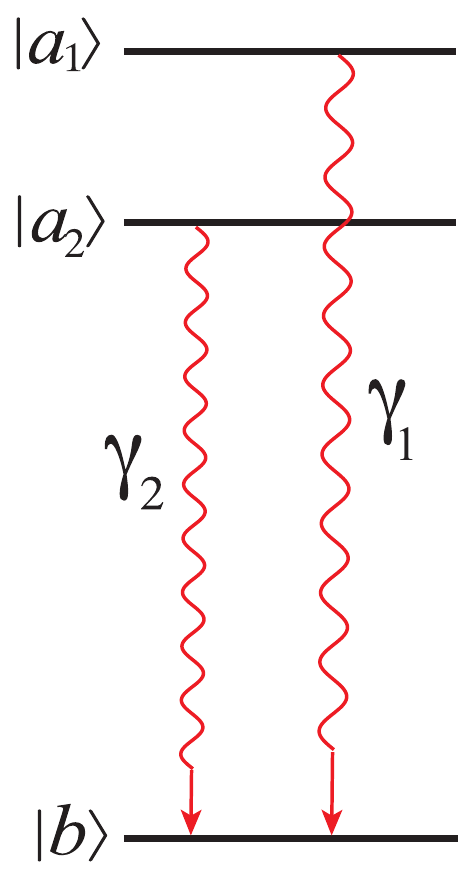}}
  \caption{Three-level atom in which the transition $|a_{1}\rangle \rightarrow |b\rangle$ and  $|a_{2}\rangle \rightarrow |b\rangle$ are coupled to the vacuum modes}\label{AFIFig1}
\end{figure}

In this section we will briefly review interaction of a three-level system with the vacuum. The three-level atom has two closely spaced upper levels as shown in Fig.(\ref{AFIFig1}). The interaction Hamiltonian in the interaction picture is 
\begin{equation}\label{D1}
\begin{split}
\mathcal{V}=\hbar\sum_{\textbf{k},\alpha}\left[g^{(a_{1}b)}_{\textbf{k},\alpha}\sigma_{+}\hat{a}^{(1)}_{\textbf{k},\alpha}e^{i\textbf{k}\cdot\textbf{R}}e^{i(\omega_{a_{1}b}-\nu_{k})t}\right.\\
\left.+g^{(a_{2}b)}_{\textbf{k},\alpha}\sigma_{+}\hat{a}^{(2)}_{\textbf{k},\alpha}e^{i\textbf{k}\cdot\textbf{R}}e^{i(\omega_{a_{2}b}-\nu_{k})t}+\text{H.c}\right]
\end{split}
\end{equation}
Redefining the coupling constant as $g_{\textbf{k},\alpha}(\textbf{R})=g_{\textbf{k},\alpha}e^{i\textbf{k}\cdot\textbf{R}}$, we write Eq. (\ref{D1}) as
\begin{equation}\label{D2}
\begin{split}
\mathcal{V}=\hbar\sum_{\textbf{k},\alpha}\left[g^{(a_{1}b)}_{\textbf{k},\alpha}(\textbf{R})\sigma_{+}\hat{a}^{(1)}_{\textbf{k},\alpha}e^{i(\omega_{a_{1}b}-\nu_{k})t}\right.\\
\left.+g^{(a_{2}b)}_{\textbf{k},\alpha}(\textbf{R})\sigma_{+}\hat{a}^{(2)}_{\textbf{k},\alpha}e^{i(\omega_{a_{2}b}-\nu_{k})t}+\text{H.c}\right]
\end{split}
\end{equation}
where the coupling constant $g^{ij}_{\textbf{k},\alpha}$ and the atomic transition operator $\sigma_{ij}$ are defined as
\begin{equation}\label{D3}
g^{ij}_{\textbf{k},\alpha}=-\frac{\wp_{ij}\cdot \textbf{e}_{\textbf{k},\alpha}\mathcal{E}_{\textbf{k},\alpha}}{\hbar}
\end{equation}
\begin{equation}\label{D4}
\sigma_{ij}=|i\rangle\langle j|
\end{equation}
Here again we will assume (without the loss of generality) that the atom in the upper level $|a_{1}\rangle$ at $t=0$ and the field modes are in the vacuum state $|\{0\}\rangle$. The state vector at $t>0$ is given by
\begin{equation}\label{D5}
\begin{split}
|\Psi(t)\rangle=C_{a_{1}}(t)|a_{1}\rangle|\{0\}\rangle +C_{a_{2}}(t)|a_{2}\rangle|\{0\}\rangle+\\ \sum_{\textbf{k},\alpha}C_{b,\textbf{k},\alpha}(t)|b\rangle|1_{\textbf{k},\alpha}\rangle
\end{split}
\end{equation}
with $C_{a_{1}}(0)=1, C_{a_{2}}=0$ and $C_{\textbf{k},\alpha}(0)=0$. The evolution equation for the state vector is governed by the Schrodinger equation $i\hbar|\dot{\Psi}(t)\rangle=\mathcal{V}|\Psi(t)\rangle$ which gives
\begin{equation}\label{D6}
\dot{C}_{a_{1}}(t)=-i\sum_{\textbf{k},\alpha}g^{(a_{1}b)}_{\textbf{k},\alpha}(\textbf{R})e^{i(\omega_{a_{1}b}-\nu_{k})t}C_{b,\textbf{k},\alpha}(t)
\end{equation}
\begin{equation}\label{D7}
\dot{C}_{a_{2}}(t)=-i\sum_{\textbf{k},\alpha}g^{(a_{2}b)}_{\textbf{k},\alpha}(\textbf{R})e^{i(\omega_{a_{2}b}-\nu_{k})t}C_{b,\textbf{k},\alpha}(t)
\end{equation}
\begin{equation}\label{D8}
\begin{split}
\dot{C}_{b,\textbf{k},\alpha}(t)=-ig^{\ast(a_{1}b)}_{\textbf{k},\alpha}(\textbf{R})e^{-i(\omega_{a_{1}b}-\nu_{k})t}C_{a_{1}}(t)\\ -ig^{\ast(a_{2}b)}_{\textbf{k},\alpha}(\textbf{R})e^{-i(\omega_{a_{2}b}-\nu_{k})t}C_{a_{1}}(t)
\end{split}
\end{equation}
Solving for $C_{b,\textbf{k},\alpha}(t)$ we obtain,
\begin{equation}\label{D9}
\begin{split}
C_{b,\textbf{k},\alpha}(t)=-ig^{\ast(a_{1}b)}_{\textbf{k},\alpha}(\textbf{R})\int_{0}^{t}e^{-i(\omega_{a_{1}b}-\nu_{k})t'}C_{a_{1}}(t')dt'\\
-ig^{\ast(a_{2}b)}_{\textbf{k},\alpha}(\textbf{R})\int_{0}^{t}e^{-i(\omega_{a_{2}b}-\nu_{k})t'}C_{a_{1}}(t')
\end{split}
\end{equation}
Formal substitution of Eq.(\ref{D9}) in Eq.(\ref{D6}) we obtain
\begin{equation}
\begin{split}
&\dot{C}_{a_{1}}(t)=-\sum_{\textbf{k},\alpha}|g^{(a_{1}b)}_{\textbf{k},\alpha}(\textbf{R})|^{2}\int_{0}^{t}e^{i(\omega_{a_{1}b}-\nu_{k})(t-t')}C_{a_{1}}(t')dt'\\
&-\sum_{\textbf{k},\alpha}g^{(a_{1}b)}_{\textbf{k},\alpha}g^{(a_{2}b)}_{\textbf{k},\alpha}\int_{0}^{t}e^{i[(\omega_{a_{1}b}-\nu_{k})t-(\omega_{a_{2}b}-\nu_{k})t']}C_{a_{2}}(t')dt'
\end{split}
\end{equation}
Using simple algebra, we can obtain
\begin{equation}
\begin{split}
\dot{C}_{a_{1}}(t)=-\frac{\gamma_{1}}{2}C_{a_{1}}(t) -p\frac{|\wp^{(a_{1}b)}||\wp^{(a_{2}b)}|}{6\pi^{2}\epsilon_{0}c^{3}\hbar }\int_{0}^{\infty}\nu^{3}_{k}d\nu_{k}\\
\times \int_{0}^{t}e^{i[(\omega_{a_{1}b}-\nu_{k})t-(\omega_{a_{2}b}-\nu_{k})t']}C_{a_{2}}(t')dt'
\end{split}
\end{equation}
where the factor $p$ is the most important dipole-orientation term defined as
\begin{equation}
p=\frac{\wp^{(a_{1}b)}\cdot \wp^{(a_{2}b)}}{|\wp^{(a_{1}b)}||\wp^{(a_{2}b)}|}
\end{equation} 
Let us consider that the splitting between the two levels $\omega_{12}=\omega_{a_{1}b}-\omega_{a_{2}b} \ll \omega_{a_{1}b},\omega_{a_{1}b}$ and using the Weiger-Weiskopf and Markov approximation approximation\cite{MOS}, we obtain 
\begin{equation}
\begin{split}
\dot{C}_{a_{1}}(t)=-\frac{\gamma_{1}}{2}C_{a_{1}}(t) -p\frac{\sqrt{\gamma_{1}\gamma_{2}}}{2}C_{a_{2}}(t)\int_{0}^{\infty}d\nu_{k}\\ \times \int_{0}^{t}e^{i[(\omega_{a_{1}}t-\omega_{a_{2}b}t')}e^{-i\nu_{k}(t-t')}]dt'
\end{split}
\end{equation}
Performing the integration over the frequency variable and using the result $\int_{0}^{\infty}e^{-i\nu_{k}(t-t')} d\nu_{k}=\delta(t-t')$ we obtain
\begin{equation}\label{D14}
\begin{split}
\dot{C}_{a_{1}}(t)=-\frac{\gamma_{1}}{2}C_{a_{1}}(t) -p\frac{\sqrt{\gamma_{1}\gamma_{2}}}{2}C_{a_{2}}(t)\\
\times \int_{0}^{\infty}\int_{0}^{\infty}e^{i[(\omega_{a_{1}}t-\omega_{a_{2}b}t')}\delta(t-t')dt'
\end{split}
\end{equation}
Here we have neglected the Lamb-shift term. Thus we arrive at the final form the Eq.(\ref{D14}) as 
\begin{equation}\label{D15}
\begin{split}
\dot{C}_{a_{1}}(t)=-\frac{\gamma_{1}}{2}C_{a_{1}}(t) -p\frac{\sqrt{\gamma_{1}\gamma_{2}}}{2}C_{a_{2}}(t)e^{i\omega_{12} t}
\end{split}
\end{equation}
The second term is counterintuitive and results from the coupling of the two transition with the vacuum modes. Similarly we can obtain the evolution equation for $\dot{C}_{a_{2}}(t)$ as 
\begin{equation}\label{D16}
\begin{split}
\dot{C}_{a_{2}}(t)=-\frac{\gamma_{1}}{2}C_{a_{2}}(t) -p\frac{\sqrt{\gamma_{1}\gamma_{2}}}{2}C_{a_{1}}(t)e^{-i\omega_{12}t}
\end{split}
\end{equation}
To eliminate the fast oscillating term $e^{i\omega_{12}t}$ we will make a transformation $C_{a_{1}}(t)=a_{1}(t)$ and $C_{a_{2}}(t)=a_{2}(t)e^{-i\omega_{12}t}$, thus Eq.(\ref{D15}),(\ref{D16}) gives
\begin{equation}\label{D17}
\begin{split}
\dot{a_{1}}(t)=-\frac{\gamma_{1}}{2}a_{1}(t) -p\frac{\sqrt{\gamma_{1}\gamma_{2}}}{2}a_{2}(t)
\end{split}
\end{equation}
\begin{equation}\label{D18}
\begin{split}
\dot{a_{2}}(t)=-\left(\frac{\gamma_{2}}{2}-i\omega_{12}\right)a_{2} -p\frac{\sqrt{\gamma_{1}\gamma_{2}}}{2}a_{1}(t)
\end{split}
\end{equation}
If we consider, degenerate upper levels then Eq.(\ref{D18}) gives
\begin{equation}\label{D19}
\begin{split}
\dot{a_{2}}(t)=-\frac{\gamma_{2}}{2}a_{2} -p\frac{\sqrt{\gamma_{1}\gamma_{2}}}{2}a_{1}(t)
\end{split}
\end{equation}
Thus even vacuum of electromagnetic field can induce non-zero coherence between degenerate levels. 
\section{Density Matrix Vs Rate Equations for TLA}
We consider a two level system with $|a\rangle$ and $|b\rangle$ as the upper and the lower levels. Population from the two levels decay out at a rate $\gamma$ while the rate of populating upper level is $r$.  The density matrix equations are given as
\begin{figure}[t]
  \centerline{\includegraphics[height=5cm,width=5cm,angle=0]{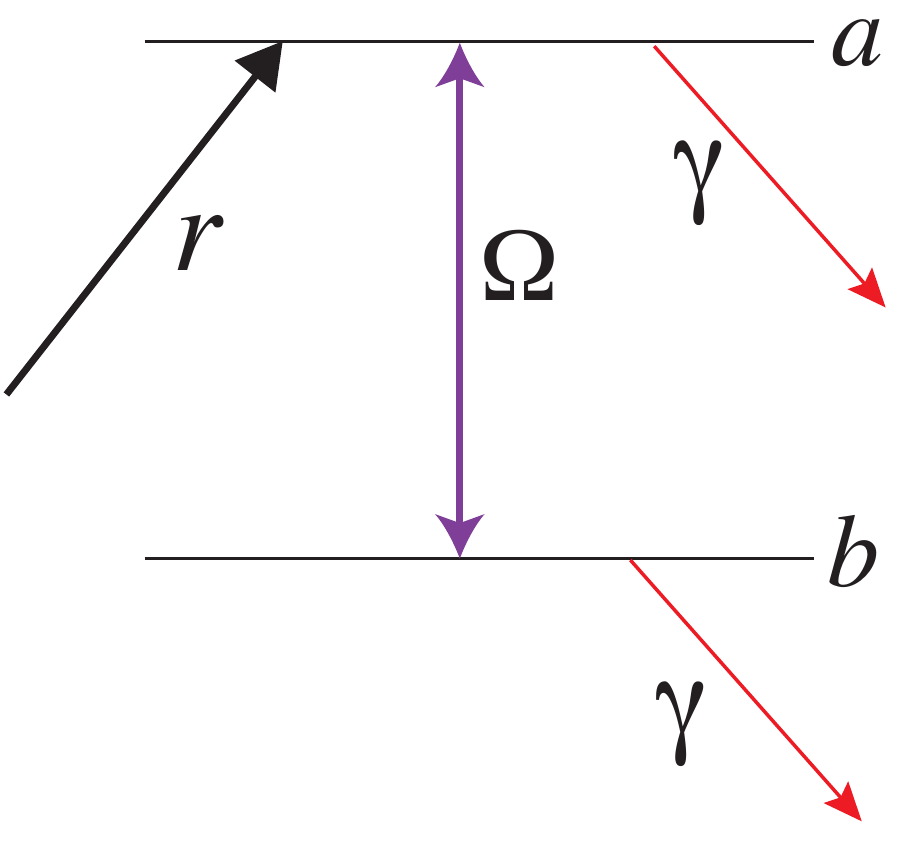}}
  \caption{Two-level model. The decay rate from the levels $|a\rangle$ and $|b\rangle$ is given by $\gamma$ while the rate of populating upper level is $r$.} 
\end{figure} 
\begin{equation}\label{E1}
\dot{\varrho}_{aa}=r-\gamma\varrho_{aa}-i\left(\Omega^{\ast}\varrho_{ab}-\Omega\varrho_{ab}^{\ast}\right)
\end{equation}
\begin{equation}\label{E2}
\dot{\varrho}_{bb}=-\gamma\varrho_{bb}+i\left(\Omega^{\ast}\varrho_{ab}-\Omega\varrho_{ab}^{\ast}\right)
\end{equation}
\begin{equation}\label{E3}
\dot{\varrho}_{ab}=-\Gamma_{ab}\varrho_{ab}-i\Omega\left(\varrho_{aa}-\varrho_{bb}\right)
\end{equation}
The propagation equation for the field $(\Omega)$ in the slowly varying amplitude approximation as
\begin{equation}\label{E4}
\frac{\partial \Omega}{\partial z}+\frac{1}{c}\frac{\partial \Omega}{\partial t}= i\eta \varrho_{ab}
\end{equation}
where the coupling constant $\eta $ is 
\begin{equation}\label{E5}
\eta = \nu N \wp^{2}/2\epsilon_{0}c\hbar
\end{equation}
Substituting $\dot{\varrho}_{ab}=0$ in Eq.(\ref{E3}) we obtain,
\begin{equation}\label{E6}
\varrho_{ab}=-i\frac{\Omega}{\Gamma_{ab}}\left(\varrho_{aa}-\varrho_{bb}\right)
\end{equation}
Substituting Eq.(\ref{E6}) in Eq.(\ref{E1}) and Eq.(\ref{E2}) we obtain,
\begin{equation}\label{E7}
\dot{\varrho}_{aa}=r-\gamma\varrho_{aa}-\frac{2\Omega^{2}}{\Gamma_{ab}}\left(\varrho_{aa}-\varrho_{bb}\right)
\end{equation}
\begin{equation}\label{E8}
\dot{\varrho}_{bb}=-\gamma\varrho_{aa}+\frac{2\Omega^{2}}{\Gamma_{ab}}\left(\varrho_{aa}-\varrho_{bb}\right)
\end{equation}
We obtain,
\begin{equation}\label{E9}
\dot{\varrho}_{aa}-\dot{\varrho}_{bb}=r-\gamma\left(\varrho_{aa}-\varrho_{bb}\right)-\frac{4\Omega^{2}}{\Gamma_{ab}}\left(\varrho_{aa}-\varrho_{bb}\right)
\end{equation}
Assuming the spatial uniformity of the field and using Eq.(\ref{E6}), the propagation equation for the field gives,
\begin{equation}\label{E10}
\frac{d \Omega}{dt}= \frac{c\eta}{\Gamma_{ab}}\left(\varrho_{aa}-\varrho_{bb}\right)\Omega
\end{equation}
\begin{figure}[t]
  \centerline{\includegraphics[height=5cm,width=8cm,angle=0]{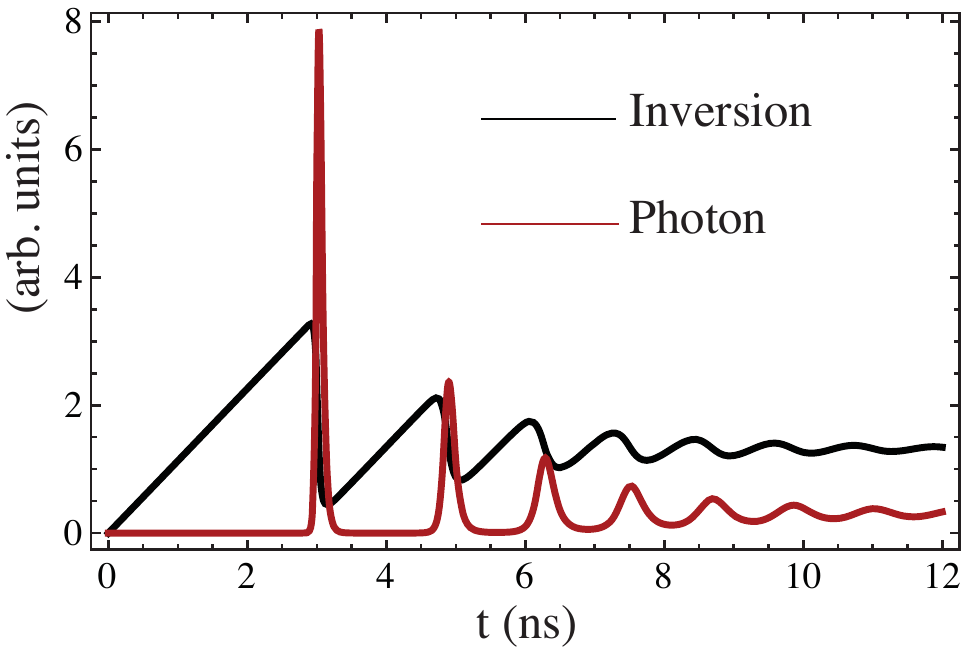}}
  \caption{Numerical simulation using the Rate equations. Using the parameters $\gamma_{c}=30$ ns$^{-1}$.} \label{RO}
\end{figure} 
Using simple algebra we obtain (for real field)
\begin{equation}\label{E11}
\frac{d \Omega^{2}}{dt}= \frac{2c\eta}{\Gamma_{ab}}\left(\varrho_{aa}-\varrho_{bb}\right)\Omega^{2}
\end{equation}
From the definition of Rabi frequency $\Omega$ and field amplitude $\mathcal{E}$ we can write,
\begin{equation}\label{E12}
\begin{split}
\Omega=\wp\mathcal{E}/2\hbar, \quad \mathcal{E}^{2}=n\hbar\nu/\epsilon_{0}V
\end{split}
\end{equation}
Here $n$ is the number of photons. From Eq.(\ref{E11}) and Eq.(\ref{E12}) we obtain,
\begin{equation}\label{E13}
\frac{d n}{dt}= \frac{\wp^{2}\nu N}{\epsilon_{0}\hbar\Gamma_{ab}V}\left(\varrho_{aa}-\varrho_{bb}\right)n
\end{equation}
From Eq.(\ref{E9}) and Eq.(\ref{E12}) we obtain,
\begin{equation}\label{E14}
\dot{\varrho}_{aa}-\dot{\varrho}_{bb}=r-\gamma\left(\varrho_{aa}-\varrho_{bb}\right)-\frac{\nu \wp^{2}}{\epsilon_{0}\hbar\Gamma_{ab}}n\left(\varrho_{aa}-\varrho_{bb}\right)
\end{equation}
Let us define some parameters to get our result in consistent with\cite{Siegman}
\begin{equation}\label{E15}
N_{ab}=NV\left(\varrho_{aa}-\varrho_{bb}\right), \quad R_{p}=NVr, \quad K=\frac{\nu \wp^{2}}{\epsilon_{0}\hbar\Gamma_{ab}V}
\end{equation}
Using new parameters, our equations takes the form
\begin{equation}\label{E16}
\frac{d n}{dt}=KN_{ab}n, \quad \dot{N}_{ab}=R_{p}-\gamma N_{ab}-KnN_{ab}
\end{equation}
To work on the numerical simulations we can use the rate equations derived from the density matrix equations for different choices of $T_{2}$. We have also added the cavity decay term in the equation of motion for $n$ phenomenologically we gives us,
\begin{equation}\label{E18}
\frac{d n}{dt}= \frac{\wp^{2}\nu N}{\epsilon_{0}\hbar\Gamma_{ab}V}\left(\varrho_{aa}-\varrho_{bb}\right)n-\gamma_{c}n
\end{equation}
In Fig.\ref{RO} we have plotted the population inversion and number of photons generated. We can clearly see the relaxation oscillations. To summarize, rate equations can be derived from the density matrix or often known as master equation when the relaxation rate of the off-diagonal term $\varrho_{ab}$ is high thus reaching steady-state value much faster than any other time scale.. Similar equations can also be easily obtained for three-level and the four-level model for traditional lasing schemes.  


\begin{thebibliography}{99}
\bibitem{Hanle}One of the first experiments to demonstrate the role of atomic coherence was reported in W. Hanle, Z. Phys. \textbf{30}, 93 (1924).
\bibitem{MOS} M. O. Scully and M. S. Zubairy, \textit{Quantum Optics}, (Cambridge University Press, Cambridge, England, 1997).
\bibitem{CPT} G. Alzetta, A. Gozzini, L. Moi and G. Orriols, Nuovo Cimento, \textbf{36B}, 5 (1976).
\bibitem{CPT1} E. Arimondo and G. Orriols, Nuovo Cimento Lett. \textbf{17}, 333 (1976).
\bibitem{CPT2} R. M. Whitley and C. R. Stroud, Jr., Phys. Rev. A \textbf{14}, 1498 (1978).
\bibitem{LWI}For review articles on LWI theory and concepts see O. Kocharovskaya, Phys. Rep. \textbf{219}, 175 (1992); S. Harris, Phys. Today \textbf{50}, 36 (1997); Ref[2].
\bibitem{LWI12} E. S. Fry, X. Li, D. Nikonov, G. G. Padmabandu, M. O. Scully, A. V. Smith, F. K. Tittel, C. Wang, S. R. Wilkinson and S. Y. Zhu, Phys. Rev. Lett. {\textbf{70}}, 3235 (1993). 
\bibitem{LWI4} A. S. Zibrov, M. D. Lukin, D. E. Nikonov, L. Hollberg, M. O. Scully, V. L. Velichansky, and H. G. Robinson,  Phys. Rev. Lett. {\textbf{75}}, 1499 (1995).
\bibitem{LWI5} G. G. Padmabandu, G. R. Welch, I. N. Shubin, E. S. Fry, D. E. Nikonov, M. D. Lukin, and M. O. Scully, Phys. Rev. Lett. {\textbf{76}}, 2053 (1996). 
\bibitem{SL} M. M. Kash, V. A. Sautenkov, A. S. Zibrov, L. Hollberg, G. R. Welch, M. D. Lukin, Y. Rostovtsev, E. S. Fry, and M. O. Scully, Phys. Rev. Lett.  \textbf{82}, 5229 (1999).
\bibitem{SL1} D. Budker, D. F. Kimball, S. M. Rochester, and V. V. Yashchuk, Phys. Rev. Lett.  \textbf{83}, 1767 (1999).
\bibitem{SL2} O. Kocharovskaya, Y. V. Rostovtsev, and M. O. Scully, Phys. Rev. Lett.  \textbf{86}, 628(2001).
\bibitem{REE} M.O. Scully, Phys. Rev. Lett. \textbf{67}, 1855 (1991).
\bibitem{REE1} U. Rathe, M. Fleischhauer, S. Y.  Zhu, T. W. Hansch, M.O. Scully, Phys. Rev. A \textbf{47}, 4994 (1993).
\bibitem{REE2} M. D. Lukin, S. F. Yelin, A. S. Zibrov, M. O. Scully, Laser Phys. \textbf{9}, 759 (1999).
\bibitem{REE3} J.P. Dowling, C.M. Bowden, Phys. Rev. Lett. \textbf{70}, 1421 (1993).
\bibitem{HSM} M. Fleischhauer and M. O. Scully, Phys. Rev. A \textbf{49},1973 (1994).
\bibitem{HSM1}  H. Lee, M. Fleischhauer and M.O. Scully, Phys. Rev. A \textbf{58}, 2587 (1998).
\bibitem{CRU}L. Yuan, A. A. Lanin, P. K. Jha, A. J. Traverso, D. V. Voronine, K. E. Dorfman, A. B. Fedotov, G. R. Welch, A. V. Sokolov, A. M. Zheltikov, and M. O. Scully, Laser Phys. Lett.  \textbf{8}, 736 (2011).
\bibitem{Jain93} M. Jain, J. E. Field, and G. Y. Yin, Opt. Lett. \textbf{18}, 998 (1993).
\bibitem{Jain96} M. Jain, H. Xia, G. Y. Yin, A. J. Merriam, and S. E. Harris, Phys. Rev. Lett. \textbf{77}, 4326 (1996).
\bibitem{Voronine12} D. V. Voronine, A. M. Sinyukov, X. Hua, K. Wang, P. K. Jha, E. Munusamy, S. E. Wheeler, G. R. Welch, A. V. Sokolov, and M. O. Scully,  Sci. Rep. \textbf{2}, 891(2012).
\bibitem{JhaAPL12} P. K. Jha,  K. E. Dorfman, Z. Yi, L. Yuan, Y. V. Rostovtsev, V. A. Sautenkov, G. R. Welch, A. M. Zheltikov, and M. O. Scully, Appl. Phys. Lett. 101, 091107 (2012).
\bibitem{JhaPQE11} P. K. Jha, H. Eleuch, and Y. V. Rostovtsev, J. Mod. Opt. 58, 652 (2011).
\bibitem{Engle07} G. S. Engel, T. R. Calhoun, E. L. Read, T. K. Ahn,T. Mancal, Y.C Cheng, R.E. Blankenship and G.R.Fleming, Nature \textbf{446} 782 (2007).
\bibitem{Collini10} E. Collini, C.Y. Wong, K. E. Wilk, P. M. Curmi, P. Brumer and G. D. Scholes Nature \textbf{463} 644(2010).
\bibitem{Pani10} G. Panitchayangkoon, D. Hayes, K. A. Fransted, J. R. Caram, E. Harel, J. Wen, R. E. Blankenship and G. S. Engel Proc. Nat. Acad. Sci. \textbf{107} 12766(2010).
\bibitem{Woj08} A. K. Wojcik, F. Xie, V. R. Chaganti, A. A. Belyanin, and J. Kono, J. Mod. Opt.  {\bf 55}, 3305 (2008).
\bibitem{Vas11} P. Vasinajindakaw, J. Vaillancourt, G. Gu, R. Liu, Y. Ling, and X. Lu, App. Phys. Lett. {\bf 98}, 211111 (2011).
\bibitem{Cap97} J. Faist, F. Capasso, C. Sirtori, K. W. West, and L. N. Pfeiffer, Nature {\bf 390}, 589 (1997).
\bibitem{Sch97} H. Schmidt, K.L. Campman, A. C. Gossard, and A. Imamoglu, Appl. Phys. Lett. \textbf{70}, 3455 (1997).
\bibitem{And10}A. E. Miroshnichenko, S. Flach, and Y. S. Kivshar, Rev. Mod. Phys. {\bf 82}, 2257 (2010).
\bibitem{Gib11} H. M. Gibbs, G. Khitrova and S. W. Koch, Nat. Phot {\bf 5}, 275 (2011).
\bibitem{Agarwal74} G. S. Agarwal, \textit{Quantum Optics}, Springer Tracts in Modern Physics Vol. 70 (Springer, Berlin, 1974), p. 95.
\bibitem{Li09} H. Li, V. A. Sautenkov, Y. V. Rostovtsev, G. R. Welch, P. R. Hemmer, and M. O. Scully, Phys. Rev. A \textbf{80}, 023820 (2009).
\bibitem{Zibrov02} A. S. Zibrov, A. B. Matsko, and M. O. Scully, Phys. Rev. Lett. \textbf{89}, 103601 (2002). 
\bibitem{JhaPRA13}P. K. Jha, S. Das and T. N. Dey (unpublished) arXiv:1210.2356
\bibitem{Carter} S. G. Carter, V. Birkedal, C. S. Wang, L. A. Coldren, A. V. Maslov, D. S. Citrin, and M. S. Sherwin, Science {\bf{310}}, 651(2005).
\bibitem{Dorfman13}K. E. Dorfman, P. K. Jha, D. V. Voronine, P. Genevet, F. Capasso and M. O. Scully, arXiv:1212.523.
\bibitem{PKJha13}P. K. Jha, X. Yin and X. Zhang, Appl. Phys. Lett. 102, 091111(2013).
\bibitem{Mavro} N. E. Mavromatos, J. Phys.: Conf. Ser. {\bf{329}} 012026 (2011).
\bibitem{Mollow69} B. R. Mollow, Phys. Rev. \textbf{188}, 1969 (1969).
\bibitem{Mollow77} F. Y. Wu, S. Ezekiel, M. Ducloy, and B. R. Mollow, Phys. Rev. Lett. \textbf{38}, 1077 (1977).
\bibitem{Valle10} E. del Valle and F. P. Laussy, Phys. Rev. Lett. \textbf{105}, 233601 (2010).
\bibitem{Scu10} M. O. Scully, Phys. Rev. Lett. \textbf{104}, 207701 (2010).
\bibitem{Scu11} M. Scully, K. Chapin, K. Dorfman, M. Kim, and A. Svidzinsky,  Proc. Natl. Acad. Sci. U.S.A \textbf{108}, 15097 (2011).
\bibitem{Jha11} K. E. Dorfman, P. K. Jha, and S. Das, Phys. Rev. A \textbf{84}, 053803 (2011).
\bibitem{mos08jmo} M.O. Scully, Y. Rostovtsev, A. Svidzinsky, J-T Chang, J. Mod. Opt. \textbf{55}, 3219 (2008).
\bibitem{Dicke54} R. Dicke, Phys. Rev.  \textbf{93}, 99, (1954).
\bibitem{TLWI2} E. A. Sete, A. A. Svidzinsky, Y. V. Rostovtsev, H. Eleuch, P. K. Jha, S. Suckewer and M. O. Scully, IEEE J. Sel. Top. Quantum Electron. \textbf{18}, 541 (2012).
\bibitem{PKJha12} P. K. Jha, A. A. Svidzinsky, and M. O. Scully, Laser Phys. Lett. \textbf{9}, 368 (2012).
\bibitem{PKJha11} P. K. Jha, J. Mod. Opt. \textbf{58}, 1957 (2011).
\bibitem{Mollow72}B. R. Mollow, Phys, Rev, A {\bf{5}} 2217 (1972). 
\bibitem{JhaCEP1} P. K. Jha, H. Li, V. A. Sautenkov, Y. V. Rostovtsev, and M. O. Scully, Opt. Commun. \textbf{284}, 2538 (2011).
\bibitem{JhaCEP2}P. K. Jha, Y. V. Rostovtsev, H. Li, V. A. Sautenkov, and M. O. Scully, Phys. Rev. A \textbf{83}, 033404 (2011).
\bibitem{JhaPJ}P. K. Jha, H. Eleuch, and Y. V. Rostovtsev, Phys. Rev A \textbf{82}, 045805 (2010).
\bibitem{JhaPQE13}P. K. Jha, H. Eleuch, and F. Grazioso (unpublished) arXiv:1302.6541
\bibitem{JhaPRA10a}P. K. Jha, and Y. V. Rostovtsev, Phys. Rev. A 81, 033827 (2010).
\bibitem{JhaPRA10b}P. K. Jha, and Y. V. Rostovtsev, Phys. Rev. A 82, 015801 (2010).
\bibitem{TLWI1} S. Ya. Kilin, K. T. Kapale and M. O. Scully, Phys. Rev. Lett. \textbf{100}, 173601 (2008).
\bibitem{Payne03}W. F. Krupke, R. J. Beach, V. K. Kanz, and S. A. Payne, Opt. Lett.  \textbf{28}, 2336 (2003).
\bibitem{Matthew} M. D. Rotondaro, Ph.D. Thesis, Air Force Institute of Technology, Ohio (1995).
\bibitem{Sky11}A. Dogariu, J.B. Michael, M.O. Scully, and R.B. Miles, Science \textbf{331}, 442 (2011).
\bibitem{Y1} A. Imamoglu and S. E. Harris, Opt. Lett. \textbf{14}, 1344 (1989).
\bibitem{Y2} A. Imamoglu, J. E. Field, and S. E. Harris, Phys. Rev. Lett. \textbf{66}, 1154 (1991).
\bibitem{Shuker1} D. Braunstein and R. Shuker, Phys. Rev. A \textbf{68}, 013812 (2003).
\bibitem{Siegman} See page 958 from A.S.Siegman, \textit{LASERS}, (University Science Books, California, USA, 1986).
\end{thebibliography}
\end{document}